# Linearized and Single-Pass Belief Propagation

Wolfgang Gatterbauer
Carnegie Mellon University
gatt@cmu.edu

Stephan Günnemann
Carnegie Mellon University
sguennem@cs.cmu.edu

Danai Koutra
Carnegie Mellon University
dkoutra@cs.cmu.edu

Christos Faloutsos
Carnegie Mellon University
christos@cs.cmu.edu

## ABSTRACT

How can we tell when accounts are fake or real in a social network? And how can we tell which accounts belong to liberal, conservative or centrist users? Often, we can answer such questions and label nodes in a network based on the labels of their neighbors and appropriate assumptions of homophily ("birds of a feather flock together") or heterophily ("opposites attract"). One of the most widely used methods for this kind of inference is *Belief Propagation (BP)* which iteratively propagates the information from a few nodes with explicit labels throughout a network until convergence. One main problem with BP, however, is that there are no known exact guarantees of convergence in graphs with loops.

This paper introduces *Linearized Belief Propagation* (LinBP), a linearization of BP that allows a closed-form solution via intuitive matrix equations and, thus, comes with convergence guarantees. It handles homophily, heterophily, and more general cases that arise in multi-class settings. Plus, it allows a compact implementation in SQL. The paper also introduces *Single-pass Belief Propagation* (SBP), a "localized" version of LinBP that propagates information across every edge at most once and for which the final class assignments depend only on the nearest labeled neighbors. In addition, SBP allows fast incremental updates in dynamic networks. Our runtime experiments show that LinBP and SBP are orders of magnitude faster than standard BP, while leading to almost identical node labels.

## 1. INTRODUCTION

Network effects are powerful and often appear in terms of *homophily* ("birds of a feather flock together"). For example, if we know the political leanings of most of Alice's friends on Facebook, then we have a good estimate of her leaning as well. Occasionally, the reverse is true, also called *heterophily* ("opposites attract"). For example, in an online dating site, we may observe that talkative people prefer to date silent ones, and vice versa. Thus, knowing the labels of a few nodes in a network, plus knowing whether homophily or heterophily applies in a given scenario, we can usually give good predictions about the labels of the remaining nodes.

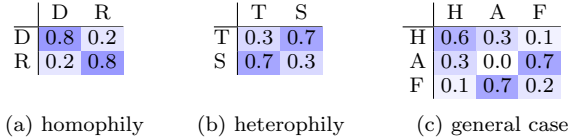

Figure 1: Three types of *network effects* with example coupling matrices. Shading intensity corresponds to the affinities or coupling strengths between classes of neighboring nodes. (a): D: Democrats, R: Republicans. (b): T: Talkative, S: Silent. (c): H: Honest, A: Accomplice, F: Fraudster.

In this work, we not only cover these two popular cases with $k=2$ classes, but also capture more general settings that mix homophily and heterophily. We illustrate with an example taken from online auction settings like e-bay [36]: Here, we observe $k=3$ classes of people: fraudsters (F), accomplices (A) and honest people (H). Honest people buy and sell from other honest people, as well as accomplices; accomplices establish a good reputation (thanks to multiple interactions with honest people), they never interact with other accomplices (waste of effort and money), but they do interact with fraudsters, forming near-bipartite cores between the two classes. Fraudsters primarily interact with accomplices (to build reputation); the interaction with honest people (to defraud them) happens in the last few days before the fraudster's account is shut down.

Thus, in general, we can have $k$ different classes, and $k \times k$ affinities or coupling strengths between them. These affinities can be organized in a coupling matrix (which we call *heterophily matrix*[1]), as shown in Fig. 1 for our three examples. Figure 1a shows the matrix for homophily: It captures that a connection between people with similar political orientations is more likely than between people with different orientations.[2] Figure 1b captures our example for heterophily: Class T is more likely to date members of class S, and vice versa. Finally, Fig. 1c shows our more general example: We see homophily between members of class H and heterophily between members of classes A and F.

In all of the above scenarios, we are interested in the most likely "beliefs" (or labels) for all nodes in the graph. The underlying problem is then: *How can we assign class labels when we know who-contacts-whom and the apriori ("explicit") labels for some of the nodes in the network?* This learning scenario, where we reason from observed training cases directly to test cases, is also called *transductive inference*, or semi-supervised learning (SSL).[3]

---
[1] In this paper, we assume the heterophily matrix to be given; e.g., by domain experts. Learning the heterophily matrix from existing (partially) labeled data is interesting future work.
[2] An example of homophily with $k=4$ classes would be co-authorship: Researchers in computer science, physics, chemistry and biology, tend to publish with co-authors of similar training. Efficient labeling in case of homophily is possible; e.g., by simple relational learners [29].

One of the most widely used methods for this kind of transductive inference in networked data is *Belief Propagation* (BP) [41], which has been successfully applied in scenarios, such as fraud detection [30, 36] and malware detection [5]. BP propagates the information from a few explicitly labeled nodes throughout the network by iteratively propagating information between neighboring nodes. However, BP has well-known convergence problems in graphs with loops (see [41] for a detailed discussion from a practitioner's point of view). While there is a lot of work on convergence of BP (e.g., [8, 32]), *exact* criteria for convergence are not known [33, Sec. 22]. In addition, whenever we get additional explicit labels (e.g., we identify more fraudsters in the online auction setting), we need to re-run BP from scratch. These issues raise fundamental theoretical questions of practical importance: *How can we find sufficient and necessary conditions for convergence of the algorithm? And how can we support fast incremental updates for dynamic networks?*

**Contributions.** This paper introduces two new formulations of BP. Unlike standard BP, these ($i$) come with exact convergence guarantees, ($ii$) allow closed-form solutions, ($iii$) give a clear intuition about the algorithms, ($iv$) can be implemented on top of standard SQL, and ($v$) one can even be updated incrementally. In more detail, we introduce:

*(1) LinBP*: Section 3 gives a new matrix formulation for multi-class BP called *Linearized Belief Propagation* (LinBP). Section 4 proves LinBP to be the result of applying a certain linearization process to the update equations of BP. Section 4.2 goes one step further and shows that the solution to LinBP can be obtained in *closed-form* by the inversion of an appropriate Kronecker product. Section 5.1 shows that this new closed-form provides us with *exact convergence guarantees* (even on graphs with loops) and a clear intuition about the reasons for convergence/non-convergence. Section 5.3 shows that our linearized matrix formulation of LinBP allows a compact implementation in SQL with standard joins and aggregates, plus iteration. Finally, experiments in Sect. 7 show that a main-memory implementation of LinBP takes 4 sec for a graph for which standard BP takes 40 min, while giving almost identical classifications ($> 99.9\%$ accuracy).

*(2) SBP*: Section 6 gives a novel semantics for "local" transductive reasoning called *Single-pass Belief Propagation* (SBP). SBP propagates information across every edge *at most once* (i.e. it ignores some edges) and is a generalization of relational learners [29] from homophily to heterophily and even more general couplings between classes in a sound and intuitive way. In particular, the final labels depend only on the nearest neighbors with explicit labels. The intuition is simple: If we do not know the political leanings of Alice's friends, than knowing the political leaning of *friends of Alice's friends* (i.e. nodes that are 2 hops away in the underlying network) will help us make some predictions about her. However, if we do know about most of her friends, then information that is more distant in the network can often be safely ignored. We formally show the connection between LinBP and SBP by proving that the labeling assignments for both are identical in the case of decreasing affinities between nodes in a graph. Importantly, SBP (in contrast to standard BP and LinBP) allows fast *incremental maintenance* for the predicated labels if the underlying network is dynamic: Our SQL implementation of SBP allows incremental updates with an intuitive index based on shortest paths to explicitly labeled nodes. Finally, experiments in Sect. 7 show that a disk-bound implementation of SBP is even faster than LinBP by one order of magnitude while giving similar classifications ($> 98.6\%$ accuracy).

**Outline.** Sect. 2 provides necessary background on BP. Sect. 3 introduces the LinBP matrix formulation. Sect. 4 sketches its derivation. Sect. 5 provides convergence guarantees, extends LinBP to weighted graphs, and gives a SQL implementation of LinBP. Sect. 6 introduces the SBP semantics, including a SQL implementation for incremental maintenance. Sect. 7 gives experiments. Sect. 8 contrasts related work, and Sect. 9 concludes. All proofs plus an additional algorithm for incrementally updating SBP when adding edges to the graph are available in our technical report on ArXiv [12]. The actual SQL implementations of LinBP and SBP are available on the authors' home pages.

## 2. BELIEF PROPAGATION

*Belief Propagation* (BP), also called the sum-product algorithm, is an exact inference method for graphical models with tree structure [38]. The idea behind BP is that all nodes receive messages from their neighbors in parallel, then they update their belief states, and finally they send new messages back out to their neighbors. In other words, at iteration $i$ of the algorithm, the posterior belief of a node $s$ is conditioned on the evidence that is $i$ steps away from $s$ in the underlying network. This process repeats until convergence and is well-understood on trees.

When applied to loopy graphs, however, BP is not guaranteed to converge to the marginal probability distribution. Indeed, Judea Pearl, who invented BP, cautioned about the indiscriminate use of BP in loopy networks, but advocated to use it as an approximation scheme [38]. More important, loopy BP is not even guaranteed to converge at all. Despite this lack of exact criteria for convergence, many papers have since shown that "loopy BP" gives very accurate results *in practice* [46], and it is thus widely used today in various applications, such as error-correcting codes [27] or stereo imaging in computer vision [9]. Our practical interest in BP comes from the fact that it is not just an efficient inference algorithm on probabilistic graphical models, but it has also been successfully used for *transductive inference*.

The transductive inference problem appears, in its generality, in a number of scenarios in both the database and machine learning communities and can be defined as follows: Consider a set of keys $X = \{x_1, \ldots, x_n\}$, a domain of values $Y = \{y_1, \ldots, y_k\}$, a partial labeling function $l : X_L \to Y$ with $X_L \subseteq X$ that maps a subset of the keys to values, a weighted mapping $w : (X_1, X_2) \to \mathbb{R}$ with $(X_1, X_2) \subseteq X \times X$, and a local condition $f_i(X, w, x_i, l_i)$ that needs to hold for a solution to be accepted.[4] The three problems are then to find: ($i$) an appropriate semantics that determines labels for all keys, (ii) an efficient algorithm that implements this semantics, and (iii) efficient ways to update labels in case the labeling function $l$, or the mapping $w$ change.

---
[3]Contrast this with *inductive learning*, where we first infer *general rules* from training cases, and only then apply them to new test cases.

[4]Notice that update equations define a local condition implicitly by giving conditions that a solution needs to fulfill after convergence.

In our scenario, we are interested in the most likely beliefs (or classes) for all nodes in a network. BP helps to iteratively propagate the information from a few nodes with explicit beliefs throughout the network. More formally, consider a graph of $n$ nodes (or keys) and $k$ possible classes (or values). Each node maintains a $k$-dimensional *belief vector* where each element $i$ represents a weight proportional to the belief that this node belongs to class $i$. We denote by $\mathbf{e}_s$ the vector of prior (or *explicit*) beliefs and $\mathbf{b}_s$ the vector of posterior (or *implicit* or *final*) beliefs at node $s$, and require that $\mathbf{e}_s$ and $\mathbf{b}_s$ are normalized to 1; i.e. $\sum_i e_s(i) = \sum_i b_s(i) = 1$.[5] Using $\mathbf{m}_{st}$ for the $k$-dimensional *message* that node $s$ sends to node $t$, we can write the BP update formulas [33, 45] for the belief vector of each node and the messages it sends w.r.t. class $i$ as:

$$b_s(i) \leftarrow \frac{1}{Z_s} e_s(i) \prod_{u \in N(s)} m_{us}(i) \quad (1)$$

$$m_{st}(i) \leftarrow \sum_j H_{st}(j,i)\, e_s(j) \prod_{u \in N(s) \setminus t} m_{us}(j) \quad (2)$$

Here, we write $Z_s$ for a normalizer that makes the elements of $\mathbf{b}_s$ sum up to 1, and $H_{st}(j,i)$ for a proportional "coupling weight" that indicates the relative influence of class $j$ of node $s$ on class $i$ of node $t$ (cf. Fig. 1).[6] We assume that the relative coupling between classes is the same in the whole graph; i.e. $H(j,i)$ is identical for all edges in the graph. We further require this coupling matrix $\mathbf{H}$ to be *doubly stochastic and symmetric*: (i) Double stochasticity is a necessary requirement for our mathematical derivation.[7] (ii) Symmetry is not required but follows from our assumption of undirected edges. For BP, the above update formulas are then repeatedly computed for each node until the values (hopefully) converge to the final beliefs.

The goal in our paper is to find the *top beliefs* for each node in the network, and to assign these beliefs to the respective nodes. That is, for each node $s$, we are interested in determining the classes with the highest values in $\mathbf{b}_s$.

PROBLEM 1 (TOP BELIEF ASSIGNMENT). *Given: (1) an undirected graph with $n$ nodes and adjacency matrix $\mathbf{A}$, where $A(s,t) \neq 0$ if the edge $s-t$ exists, (2) a symmetric, doubly stochastic coupling matrix $\mathbf{H}$ representing $k$ classes, where $H(j,i)$ indicates the relative influence of class $j$ of a node on class $i$ of its neighbor, and (3) a matrix of explicit beliefs $\mathbf{E}$, where $E(s,i) \neq 0$ is the strength of belief in class $i$ by node $s$. The goal of* top belief assignment *is to infer for each node a set of classes with highest final belief.*

In other words, our problem is to find a mapping from nodes to *sets of classes* (in order to allow for ties).

## 3. LINEARIZED BELIEF PROPAGATION

|  | Formula | Maclaurin series | Approx. |
|---|---|---|---|
| Logarithm | $\ln(1+\epsilon) = \epsilon - \frac{\epsilon^2}{2} + \frac{\epsilon^3}{3} - \ldots$ | | $\approx \epsilon$ |
| Division | $\frac{\frac{1}{k}+\epsilon_1}{1+\epsilon_2}$ | $= (\frac{1}{k}+\epsilon_1)(1-\epsilon_2+\epsilon_2^2-\ldots)$ | $\approx \frac{1}{k}+\epsilon_1 - \frac{\epsilon_2}{k}$ |

Figure 2: Two linearizing approximations used in our derivation.

In this section, we introduce *Linearized Belief Propagation (LinBP)*, which is a closed-form description for the final beliefs after convergence of BP under mild restrictions of our parameters. The main idea is to *center* values around default values (using Maclaurin series expansions) and to then restrict our parameters to small deviations from these defaults. The resulting equations replace multiplication with addition and can thus be put into a matrix framework with a closed-form solution. This allows us to later give exact convergence criteria based on problem parameters.

DEFINITION 2 (CENTERING). *We call a vector or matrix $\mathbf{x}$ "centered around $c$" if all its entries are close to $c$ and their average is exactly $c$.*

DEFINITION 3 (RESIDUAL VECTOR/MATRIX). *If a vector $\mathbf{x}$ is centered around $c$, then the residual vector around $c$ is defined as $\hat{\mathbf{x}} = [x_1 - c, x_2 - c, \ldots]$. Accordingly, we denote a matrix $\hat{\mathbf{X}}$ as a residual matrix if each column and row vector corresponds to a residual vector.*

For example, we call the vector $\mathbf{x} = [1.01, 1.02, 0.97]$ centered around $c = 1$.[8] The residuals from $c$ will form the *residual vector* $\hat{\mathbf{x}} = [0.01, 0.02, -0.03]$. Notice that the entries in a residual vector always sum up to 0, by construction.

The main ideas in our proofs are as follows: (1) the $k$-dimensional message vectors $\mathbf{m}$ are centered around 1; (2) all the other $k$-dimensional vectors are probability vectors, they have to sum up to 1, and thus they are centered around $1/k$. This holds for the belief vectors $\mathbf{b}$, $\mathbf{e}$, and for the all entries of matrix $\mathbf{H}$; and (3) we make use of each of the two linearizing approximations shown in Fig. 2 exactly once.

According to aspect (1) of the previous paragraph, we require that the messages sent are normalized so that the average value of the elements of a message vector is 1 or, equivalently, that the elements sum up to $k$:

$$m_{st}(i) \leftarrow \frac{1}{Z_{st}} \sum_j H(j,i)\, e_s(j) \prod_{u \in N(s) \setminus t} m_{us}(j) \quad (3)$$

Here, we write $Z_{st}$ as a normalizer that makes the elements of $\mathbf{m}_{st}$ sum up to $k$. Scaling all elements of a message vector by the same constant does *not* affect the resulting beliefs since the normalizer in Eq. 1 makes sure that the beliefs are always normalized to 1, independent of the scaling of the messages. Thus, scaling messages still preserves the exact solution, yet it will be essential for our derivation.

THEOREM 4 (LINEARIZED BP (LINBP)). *Let $\hat{\mathbf{B}}$ and $\hat{\mathbf{E}}$ be the residual matrices of final and explicit beliefs centered around $1/k$, $\hat{\mathbf{H}}$ the residual coupling matrix centered around $1/k$, $\mathbf{A}$ the adjacency matrix, and $\mathbf{D} = \text{diag}(\mathbf{d})$ the diagonal*

---

[5] Notice that here and in the rest of this paper, we write $\sum_i$ as short form for $\sum_{i \in [k]}$ whenever $k$ is clear from the context.

[6] We chose the symbol $H$ for the coupling weights as reminder of our motivating concepts of homophily and heterophily. Concretely, if $H(i,i) > H(j,i)$ for $j \neq i$, we say homophily is present, otherwise heterophily or a mix between the two.

[7] Notice that single-stochasticity could easily be constructed by taking any set of vectors of relative coupling strengths between neighboring classes, normalizing them to 1, and arranging them in a matrix.

[8] All vectors $\mathbf{x}$ in this paper are assumed to be *column vectors* $[x_1, x_2, \ldots]^\intercal$ even if written as row vectors $[x_1, x_2, \ldots]$.

Figure 3: LinBP equation (Eq. 4): Notice our matrix conventions: $\hat{H}(j,i)$ indicates the relative influence of class $j$ of a node on class $i$ of its neighbor, $A(s,t) = A(t,s) \neq 0$ if the edge $s - t$ exists, and $\hat{B}(s,i)$ is the belief in class $i$ by node $s$.

*degree matrix.* Then, the final belief assignment from belief propagation is approximated by the equation system:

$$\hat{\mathbf{B}} = \hat{\mathbf{E}} + \mathbf{A}\hat{\mathbf{B}}\hat{\mathbf{H}} - \mathbf{D}\hat{\mathbf{B}}\hat{\mathbf{H}}^2 \qquad (LinBP) \quad (4)$$

Figure 3 illustrates Eq. 4 and shows our matrix conventions. We refer to the term $\mathbf{D}\hat{\mathbf{B}}\hat{\mathbf{H}}^2$ as "echo cancellation".[9] For increasingly small residuals, the echo cancellation becomes increasingly negligible, and by further ignoring it, Eq. 4 can be further simplified to

$$\hat{\mathbf{B}} = \hat{\mathbf{E}} + \mathbf{A}\hat{\mathbf{B}}\hat{\mathbf{H}} \qquad (LinBP^*) \quad (5)$$

We will refer to Eq. 4 (with echo cancellation) as LinBP and Eq. 5 (without echo cancellation) as LinBP$^*$.

**Iterative updates.** Notice that while these equations give an implicit definition of the final beliefs after convergence, they can also be used as iterative update equations, allowing an iterative calculation of the final beliefs. Starting with an arbitrary initialization of $\hat{\mathbf{B}}$ (e.g., all values zero), we repeatedly compute the right hand side of the equations and update the values of $\hat{\mathbf{B}}$ until the process converges:

$$\hat{\mathbf{B}}_{(l+1)} \leftarrow \hat{\mathbf{E}} + \mathbf{A}\hat{\mathbf{B}}_{(l)}\hat{\mathbf{H}} - \mathbf{D}\hat{\mathbf{B}}_{(l)}\hat{\mathbf{H}}^2 \qquad (LinBP) \quad (6)$$

$$\hat{\mathbf{B}}_{(l+1)} \leftarrow \hat{\mathbf{E}} + \mathbf{A}\hat{\mathbf{B}}_{(l)}\hat{\mathbf{H}} \qquad (LinBP^*) \quad (7)$$

Thus, the final beliefs of each node can be computed via elegant matrix operations and optimized solvers, while the implicit form gives us guarantees for the convergence of this process, as explained in Sect. 5.1. Also notice that our update equations calculate beliefs directly (i.e. without having to calculate messages first); this will give us significant performance improvements as our experiments will later show.

## 4. DERIVATION OF LINBP

This section sketches the proofs of our first technical contribution: Section 4.1 linearizes the update equations of BP by centering around appropriate defaults and using the approximations from Fig. 2 (Lemma 5), and then expressesing the steady state messages in terms of beliefs (Lemma 6). Sect. 4.2 gives an additional closed-form expression for the steady-state beliefs (Proposition 7).

### 4.1 Centering Belief Propagation

We derive our formalism by centering the elements of the coupling matrix and all message and belief vectors around their natural default values; i.e. the elements of $\mathbf{m}$ around 1, and the elements of $\mathbf{H}$, $\mathbf{e}$, and $\mathbf{b}$ around $\frac{1}{k}$. We are interested in the residual values given by: $m(i) = 1 + \hat{m}(i)$, $H(j,i) = \frac{1}{k} + \hat{H}(j,i)$, $e(i) = \frac{1}{k} + \hat{e}(i)$, and $b(i) = \frac{1}{k} + \hat{b}(i)$.[10] As a consequence, $\hat{\mathbf{H}} \in \mathbb{R}^{k \times k}$ is the *residual coupling matrix* that makes explicit the relative attraction and repulsion: The sign of $\hat{H}(j,i)$ tells us if the class $j$ attracts or repels class $i$ in a neighbor, and the magnitude of $\hat{H}(j,i)$ indicates the strength. Subsequently, this centering allows us to rewrite belief propagation in terms of the residuals.

LEMMA 5 (CENTERED BP). *By centering the coupling matrix, beliefs and messages, the equations for belief propagation can be approximated by:*

$$\hat{b}_s(i) \leftarrow \hat{e}_s(i) + \frac{1}{k} \sum_{u \in N(s)} \hat{m}_{us}(i) \qquad (8)$$

$$\hat{m}_{st}(i) \leftarrow k \sum_j \hat{H}(j,i)\hat{b}_s(j) - \sum_j \hat{H}(j,i)\hat{m}_{ts}(j) \qquad (9)$$

Using Lemma 5, we can derive a closed-form description of the steady-state of belief propagation.

LEMMA 6 (STEADY STATE MESSAGES). *For small deltas of all values from their expected values, and after convergence of belief propagation, message propagation can be expressed in terms of the steady beliefs as:*

$$\hat{\mathbf{m}}_{st} = k(\mathbf{I}_k - \hat{\mathbf{H}}^2)^{-1}\hat{\mathbf{H}}(\hat{\mathbf{b}}_s - \hat{\mathbf{H}}\hat{\mathbf{b}}_t) \qquad (10)$$

*where $\mathbf{I}_k$ is the identity matrix of size $k$.*

From Lemma 6, we can finally prove Theorem 4.

### 4.2 Closed-form solution for LinBP

In practice, we will solve Eq. 4 and Eq. 5 via an iterative computation (see end of Sect. 3). However, we next give a *closed-form* solution, which allows us later to study the convergence of the iterative updates. We need to introduce two new notions: Let $\mathbf{X}$ and $\mathbf{Y}$ be matrices of order $m \times n$ and $p \times q$, respectively, and let $\mathbf{x}_j$ denote the $j$-th column of matrix $\mathbf{X}$; i.e. $\mathbf{X} = \{x_{ij}\} = [\mathbf{x}_1 \dots \mathbf{x}_n]$. First, the *vectorization* of matrix $\mathbf{X}$ stacks the columns of a matrix one underneath the other to form a single column vector; i.e.

$$\text{vec}(\mathbf{X}) = \begin{bmatrix} \mathbf{x}_1 \\ \vdots \\ \mathbf{x}_n \end{bmatrix}$$

Second, the *Kronecker product* of $\mathbf{X}$ and $\mathbf{Y}$ is the $mp \times nq$ matrix defined by

$$\mathbf{X} \otimes \mathbf{Y} = \begin{bmatrix} x_{11}\mathbf{Y} & x_{12}\mathbf{Y} & \dots & x_{1n}\mathbf{Y} \\ x_{21}\mathbf{Y} & x_{22}\mathbf{Y} & \dots & x_{2n}\mathbf{Y} \\ \vdots & \vdots & \ddots & \vdots \\ x_{m1}\mathbf{Y} & x_{m2}\mathbf{Y} & \dots & x_{mn}\mathbf{Y} \end{bmatrix}$$

---

[9] Notice that the original BP update equations send a message across an edge that excludes information received across the same edge from the other direction ("$u \in N(s) \setminus t$" in Eq. 2). In a probabilistic scenario on tree-based graphs, this term is required for correctness. In loopy graphs (without well-justified semantics), this term still compensates for two neighboring nodes building up each other's scores.

[10] Notice that we call these default values "natural" as our results imply that if we start with centered messages around 1 and set $\frac{1}{Z_{st}} = k$, then the derived messages with Eq. 3 remain centered around 1 *for any iteration*. Also notice that multiplying with a message vector with all entries 1 does not change anything. Similarly, a prior belief vector with all entries $\frac{1}{k}$ gives equal weight to each class. Finally, notice that we call "*nodes with explicit beliefs*", those nodes for which the residuals have non-zero elements ($\hat{\mathbf{e}} \neq \mathbf{0}_k$); i.e. the explicit beliefs deviate from the center $\frac{1}{k}$.

PROPOSITION 7 (CLOSED-FORM LinBP). *The closed-form solution for LinBP (Eq. 4) is given by:*

$$\boxed{\text{vec}(\hat{\mathbf{B}}) = (\mathbf{I}_{nk} - \hat{\mathbf{H}} \otimes \mathbf{A} + \hat{\mathbf{H}}^2 \otimes \mathbf{D})^{-1} \text{vec}(\hat{\mathbf{E}})} \quad (LinBP) \quad (11)$$

By further ignoring the echo cancellation $\hat{\mathbf{H}}^2 \otimes \mathbf{D}$, we get the closed-form for LinBP* (Eq. 5) as:

$$\boxed{\text{vec}(\hat{\mathbf{B}}) = (\mathbf{I}_{nk} - \hat{\mathbf{H}} \otimes \mathbf{A})^{-1} \text{vec}(\hat{\mathbf{E}})} \quad (LinBP^*) \quad (12)$$

Thus, by using Eq. 11 or Eq. 12, we are able to compute the final beliefs in a closed-form, as long as the inverse of the matrix exists. In the next section, we show the relation of the closed-form to our original update equation Eq. 6 and give exact convergence criteria.

## 5. ADDITIONAL BENEFITS OF LINBP

In this section, we give *sufficient and necessary* convergence criteria for LinBP and LinBP*, we show how our formalism generalizes to weighted graphs, and we show how our update equations can be implemented in standard SQL.

### 5.1 Update equations and Convergence

Equation 11 and Eq. 12 give us a closed-form for the final beliefs after convergence. From the Jacobi method for solving linear systems [40], we know that the solution for $\mathbf{y} = (\mathbf{I} - \mathbf{M})^{-1}\mathbf{x}$ can be calculated, under certain conditions, via the iterative update equation

$$\mathbf{y}_{(l+1)} \leftarrow \mathbf{x} + \mathbf{M}\mathbf{y}_{(l)} \quad (13)$$

These updates are known to converge for any choice of initial values for $\mathbf{y}_{(0)}$, as long as $\mathbf{M}$ has a spectral radius $\rho(\mathbf{M}) < 1$.[11] Thus, the same convergence guarantees carry over when Eq. 11 and Eq. 12 are written, respectively, as

$$\text{vec}(\hat{\mathbf{B}}_{(l+1)}) \leftarrow \text{vec}(\hat{\mathbf{E}}) + (\hat{\mathbf{H}} \otimes \mathbf{A} - \hat{\mathbf{H}}^2 \otimes \mathbf{D})\text{vec}(\hat{\mathbf{B}}_{(l)}) \quad (14)$$

$$\text{vec}(\hat{\mathbf{B}}_{(l+1)}) \leftarrow \text{vec}(\hat{\mathbf{E}}) + (\hat{\mathbf{H}} \otimes \mathbf{A})\text{vec}(\hat{\mathbf{B}}_{(l)}) \quad (15)$$

Furthermore, it follows from Proposition 7, that update Eq. 14 is equivalent to our original update Eq. 6, and thus both have the same convergence guarantees.

We are now ready to give a sufficient *and* necessary criteria for convergence of the iterative LinBP and LinBP* update equations.

LEMMA 8 (EXACT CONVERGENCE). *Necessary and sufficient criteria for convergence of LinBP and LinBP* are:*

$$LinBP \text{ converges } \Leftrightarrow \rho(\hat{\mathbf{H}} \otimes \mathbf{A} - \hat{\mathbf{H}}^2 \otimes \mathbf{D}) < 1 \quad (16)$$

$$LinBP^* \text{ converges } \Leftrightarrow \rho(\hat{\mathbf{H}}) < \frac{1}{\rho(\mathbf{A})} \quad (17)$$

In practice, computation of the largest eigenvalues can be expensive. Instead, we can exploit the fact that any norm $||\mathbf{X}||$ gives an upper bounds to the spectral radius of a matrix $\mathbf{X}$ to establish sufficient (but not necessary) and easier-to-compute conditions for convergence.

LEMMA 9 (SUFFICIENT CONVERGENCE). *Let $||\cdot||$ stand for any sub-multiplicative norm of the enclosed matrix. Then, the following are sufficient criteria for convergence:*

$$LinBP \text{ converges } \Leftarrow ||\hat{\mathbf{H}}|| < \frac{\sqrt{||\mathbf{A}||^2 + 4||\mathbf{D}||} - ||\mathbf{A}||}{2||\mathbf{D}||} \quad (18)$$

$$LinBP^* \text{ converges } \Leftarrow ||\hat{\mathbf{H}}|| < \frac{1}{||\mathbf{A}||} \quad (19)$$

*Further, let $M$ be a set of such norms and let $||\mathbf{X}||_M := \min_{||\cdot||_i \in M} ||\mathbf{X}||_i$. Then, by replacing each $||\cdot||$ with $||\cdot||_M$, we get better bounds.*

Vector/Elementwise $p$-norms for $p \in [1, 2]$ (e.g., the Frobenius norm) and all induced $p$-norms are sub-multiplicative.[12] Furthermore, vector $p$-norms are monotonically decreasing for increasing $p$, and thus: $\rho(\mathbf{X}) \leq ||\mathbf{X}||_2 \leq ||\mathbf{X}||_1$. We thus suggest using the following set $M$ of three norms which are all fast to calculate: ($i$) Frobenius norm, ($ii$) induced-1 norm, and ($iii$) induced-$\infty$ norm.

### 5.2 Weighted graphs

Note that Theorem 4 can be generalized to allow weighted graphs by simply using a weighted adjacency matrix $\mathbf{A}$ with elements $A(i, j) = w > 0$ if the edge $j - i$ exists with weight $w$, and $A(i, j) = 0$ otherwise. Our derivation remains the same, we only need to make sure that the degree $d_s$ of a node $s$ is the sum of the squared weights to its neighbors (recall that the echo cancellation goes back and forth). The weight on an edge simply scales the coupling strengths between two neighbors, and we have to add up parallel paths. Thus, Theorem 4 can be applied for *weighted* graphs as well.

### 5.3 LinBP in SQL

Much of today's data is stored in relational DBMSs. We next give a compact translation of our linearized matrix formulation into a simple implementation in SQL with standard joins and aggregates, plus iteration. As a consequence, any standard DBMS is able to perform LinBP on networked data stored in relations. An implementation of the original BP would require either a non-standard product aggregate function (with the practical side effect of often producing underflows) or the use of an additional logarithmic function. Issues with convergence would still apply [41].

In the following, we use Datalog notation extended with aggregates in the tradition of [7]. Such an aggregate query has the form $Q(\bar{x}, \alpha(\bar{y})) \coloneq C(\bar{z})$ with $C$ being a conjunction of non-negated relational atoms and comparisons, and $\alpha(\bar{y})$ being the aggregate term.[13] When translating into SQL, the head of the query $(\bar{x}, \alpha(\bar{y}))$ defines the SELECT clause, and the variables $\bar{x}$ appear in the GROUP BY clause of the query.

We use table $A(\underline{s, t}, w)$ to represent the adjacency matrix $\mathbf{A}$ with $s$ and $t$ standing for source and target node, respectively, and $w$ for weight; $E(\underline{v, c}, b)$ and $B(\underline{v, c}, b)$ to represent the explicit beliefs $\hat{\mathbf{E}}$ and final beliefs $\hat{\mathbf{B}}$, respectively, with $v$ standing for node, $c$ for class and $b$ for belief; and $H(\underline{c_1, c_2}, h)$ to represent the coupling matrix $\hat{\mathbf{H}}$ with coupling strength $h$ from a class $c_1$ on it's neighbor's class $c_2$. From these

---

[11]The spectral radius $\rho(\cdot)$ is the supremum among the absolute values of the eigenvalues of the enclosed matrix.

[12]Vector $p$-norms are defined as $||\mathbf{X}||_p = \left(\sum_i \sum_j |X(i,j)|^p\right)^{1/p}$. Induced $p$-norms, for $p = 1$ and $p = \infty$, are defined $||\mathbf{X}||_1 = \max_j \sum_i |X(i,j)|$ and $||\mathbf{X}||_\infty = \max_i \sum_j |X(i,j)|$, i.e. as maximum absolute column sum or maximum absolute row sum, respectively.

[13]Note that in a slight abuse of notation (and for the sake of conciseness), we use variables to express both attribute names and join variables in Datalog notation.

**Algorithm 1:** (LinBP) Returns the final beliefs $B$ with LinBP for a weighted network $A$ with explicit beliefs $E$, coupling strengths $H$, and calculated tables $D$ and $H_2$.

---

**Input:** $A(s,t,w), E(v,c,b), H(c_1,c_2,h), D(v,d), H_2(c_1,c_2,h)$
**Output:** $B(v,c,b)$

1 Initialize final beliefs for nodes with explicit beliefs:
  $B(s,c,b) :\!- E(s,c,b)$
2 **for** $i \leftarrow 1$ **to** $l$ **do**
3     Create two temporary views:
  $V_1(t,c_2,sum(w \cdot b \cdot h)) :\!- A(s,t,w), B(s,c_1,b), H(c_1,c_2,h)$
  $V_2(s,c_2,sum(d \cdot b \cdot h)) :\!- D(v,d), B(s,c_1,b), H_2(c_1,c_2,h)$
4     Update final beliefs:
  $B(v,c,b_1+b_2-b_3) :\!- E(v,c,b_1), V_1(v,c,b_2), V_2(v,c,b_3)$
  **return** $B(v,c,b)$

---

data, we calculate an additional table $D(\underline{v},d)$ representing the degree matrix $\mathbf{D}$, defined to allow weighted edges:[14]

$$D(s, sum(w*w)) :\!- A(s,t,w)$$

and an additional table $H_2(\underline{c_1,c_2},h)$ representing $\hat{\mathbf{H}}^2$:

$$H_2(c_1,c_2,sum(h_1 \cdot h_2)) :\!- H(c_1,c_3,h_1), H(c_3,c_2,h_2) \quad (20)$$

Using these tables, Algorithm 1 shows the translation of the update equations for LinBP into the relational model: We initialize the final beliefs with the explicit beliefs (line 1). We then create two temporary tables, $V_1(\underline{v,c},b)$ representing the result of $\mathbf{A}\hat{\mathbf{B}}\hat{\mathbf{H}}$ and $V_2(\underline{v,c},b)$ for $\mathbf{D}\hat{\mathbf{B}}\hat{\mathbf{H}}^2$ (line 3). These views are then combined with the explicit beliefs to update the final beliefs (line 4).[15] This is repeated a fixed number $l$ of times or until the maximum change of a belief between two iterations is smaller than a threshold. Finally, we return the top beliefs for each node.

COROLLARY 10 (LINBP IN SQL). *The iterative updates for LinBP can be expressed in standard SQL with iteration.*

## 6. SINGLE-PASS BELIEF PROPAGATION

Our ultimate goal with belief propagation is to assign the most likely class(es) to each unlabeled node (i.e. each node without explicit beliefs). Here, we define a semantics for top belief assignment that is closely related to BP and LinBP (it gives the same classification for increasingly small coupling weights), but that has two algorithmic advantages: (*i*) calculating the final beliefs requires to visit every node only once (and to propagate values across an edge *at most* once); and (*ii*) the beliefs can be maintained incrementally when new explicit beliefs or edges are added to the graph.

### 6.1 Scaling Beliefs

We start with a simple definition that helps us separate the relative strength of beliefs from their absolute values.

DEFINITION 11 (STANDARDIZATION). *Given a vector* $\mathbf{x} = [x_1, x_2, \ldots, x_k]$ *with* $\mu(\mathbf{x})$ *and* $\sigma(\mathbf{x})$ *being the mean and the standard deviation of the elements of* $\mathbf{x}$, *respectively. The* standardization *of* $\mathbf{x}$ *is the new vector* $\mathbf{x}' = \zeta(\mathbf{x})$ *with* $x'_i = \frac{x_i - \mu(\mathbf{x})}{\sigma(\mathbf{x})}$ *if* $\sigma \neq 0$, *and with* $x'_i = 0$ *if* $\sigma = 0$.[16]

For example, $\zeta([1,0]) = [1,-1]$, $\zeta([1,1,1]) = [0,0,0]$, and $\zeta([1,0,0,0,0]) = [2,-0.5,-0.5,-0.5,-0.5]$. The *standardized belief assignment* $\hat{\mathbf{b}}'_s$ for a node $s$ is then the standardization of the final belief assignment: $\hat{\mathbf{b}}'_s = \zeta(\hat{\mathbf{b}}_s)$. For example, assume two nodes $s$ and $t$ with final beliefs $\hat{\mathbf{b}}_s = [4,-1,-1,-1,-1]$ and $\hat{\mathbf{b}}_t = [40,-10,-10,-10,-10]$, respectively. The standardized belief assignment is then the same for both nodes: $\hat{\mathbf{b}}'_s = \hat{\mathbf{b}}'_t = [2,-0.5,-0.5,-0.5,-0.5]$ whereas the standard deviations indicate the magnitude of differences: $\sigma(\hat{\mathbf{b}}_s) = 2$, $\sigma(\hat{\mathbf{b}}_s) = 20$.

LEMMA 12 (SCALING $\hat{\mathbf{E}}$). *Scaling the explicit beliefs with a constant factor $\lambda$ leads to scaled final beliefs by $\lambda$. In other words,* $\forall \lambda \in \mathbb{R} : (\hat{\mathbf{E}} \leftarrow \lambda \cdot \hat{\mathbf{E}}) \Rightarrow (\hat{\mathbf{B}} \leftarrow \lambda \cdot \hat{\mathbf{B}})$.

PROOF. This follows immediately from Eq. 11. □

COROLLARY 13 (SCALING $\hat{\mathbf{E}}$). *Scaling $\hat{\mathbf{E}}$ with a constant factor does not change the standardized belief assignment $\hat{\mathbf{B}}'$.*

The last corollary implies that scaling the explicit beliefs has *no effect* on the top belief assignment, and thus the ultimate classification by LinBP.

### 6.2 Scaling Coupling Strengths

While scaling $\hat{\mathbf{E}}$ has no effect on the standardized beliefs, the scale of the residual coupling matrix $\hat{\mathbf{H}}$ is important. To separate (*i*) the *relative difference* among beliefs from (*ii*) their *absolute scale*, we introduce a positive parameter $\epsilon_H$ and define with $\hat{\mathbf{H}}_o$ the unscaled ("original") residual coupling matrix implicitly by: $\hat{\mathbf{H}} = \epsilon_H \hat{\mathbf{H}}_o$. This separation allows us to keep the relative scaling fixed as $\hat{\mathbf{H}}_o$ and to thus analyze the influence of the absolute scaling on the standardized belief assignment (and thereby the top belief assignment) by varying $\epsilon_H$ only.

It was previously observed in experiments [25] that the top belief assignment is the same for a large range of $\epsilon_H$ in belief propagation with binary classes, but that it deviates for very small $\epsilon_H$. Here we show that the standardized belief assignment for LinBP converges for $\epsilon_H \to 0^+$, and that any deviations are due to limited computational precision. We also give a new closed-form for the predictions of LinBP in the limit of $\epsilon_H \to 0^+$ and name this semantics *Single-Pass Belief Propagation (SBP)*. SBP has several advantages: (*i*) it is faster to calculate (we chose its name since information is propagated across each edge at most once), (*ii*) it can be maintained incrementally, and (*iii*) it provides a simple intuition about its behavior and an interesting connection to relational learners [29]. For that, we need one more notion:

DEFINITION 14 (GEODESIC NUMBER $g$). *The geodesic number $g_t$ of a node $t$ is the length of the shortest path to any node with explicit beliefs.*

Notice that any node with explicit beliefs has geodesic number 0. For the following definition, let the weight $w$ of a path $p$ be the product of the weights of its edges (if the graph is unweighted, than the weights are 1).

DEFINITION 15 (SINGLE-PASS BP (SBP)). *Given a node $t$ with geodesic number $k$, let $P_t^k$ be the set of all paths with length $k$ from a node with explicit beliefs to $t$. For any such path $p \in P_t^k$, let $w_p$ be its weight, and $\hat{\mathbf{e}}_p$ the explicit beliefs*

---

[14] Remember from Sect. 5.2 that the degree of a node in a weighted graph is the sum of the squares of the weights to all neighbors.
[15] In practice, we use union all, followed by a grouping on $v, c$.
[16] We use the symbol $\zeta$ since standardized vector elements are also varyingly called *standard scores*, z-scores, or z-values.

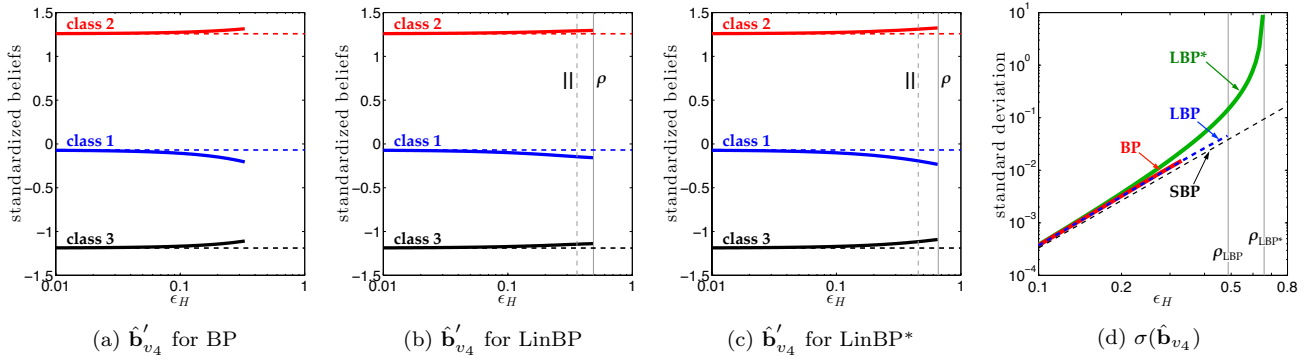

(a) $\hat{\mathbf{b}}'_{v_4}$ for BP  (b) $\hat{\mathbf{b}}'_{v_4}$ for LinBP  (c) $\hat{\mathbf{b}}'_{v_4}$ for LinBP*  (d) $\sigma(\hat{\mathbf{b}}_{v_4})$

Figure 4: Example 20: (a-c): For decreasing $\epsilon_H$, the standardized beliefs of BP, LinBP, and LinBP* converge towards the ones from SBP: $[-0.069, 1.258, -1.189]$ (horizontal dashed lines). While there are no known *exact* convergence criteria for BP, we gave *necessary and sufficient* criteria for both LinBP and LinBP* (vertical full lines named $\rho$) plus easier-to-calculate sufficient only conditions (vertical dashed lines named $||$). Notice that our $\rho$-criteria predict exactly when they algorithms stop converging (end of lines). (d): For decreasing $\epsilon_H$, the standard deviations of final beliefs for BP, LinBP, and LinBP* also converge towards the one of SBP.

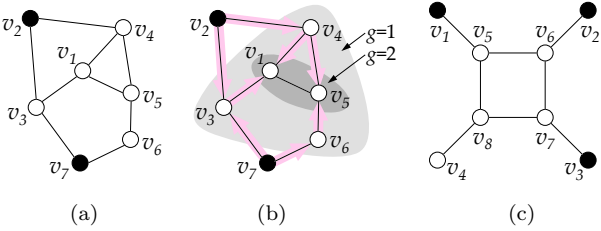

(a)  (b)  (c)

Figure 5: (a),(b): Example 16: Node $v_1$ has geodesic number 2 and three shortest paths to nodes with explicit beliefs $v_2$ and $v_7$. (c): Example 20: Example torus graph taken from [45].

of the node at the start of path $p$. The final belief assignment $\hat{\mathbf{b}}_t$ for Single-pass Belief Propagation (SBP) is defined by

$$\hat{\mathbf{b}}_s = \hat{\mathbf{H}}^k \sum_{p \in P_s^k} w_p \hat{\mathbf{e}}_p \qquad (21)$$

The intuition behind SBP is that nodes with increasing distance have an increasingly negligible influence: For every additional edge in a path, the original influence is scaled by $\epsilon_H$ times the modulation by $\hat{\mathbf{H}}$. Thus in the limit of $\epsilon_H \to 0^+$, the nearest neighbors with explicit beliefs will dominate the influence of any other node. Since linear scaling does not change the standardization of a vector, $\zeta(\epsilon \mathbf{x}) = \zeta(\mathbf{x})$, scaling $\hat{\mathbf{H}}$ has no effect on the standardized and thus also top belief assignments for SBP. In other words, the standardized belief assignment of SBP is independent of $\epsilon_H$ (as long as $\epsilon_H > 0$), and w.l.o.g. we can therefore use the unscaled coupling matrix $\hat{\mathbf{H}}_o$ ($\epsilon_H = 1$). This does not hold for LinBP.

EXAMPLE 16 (SBP ILLUSTRATION). *Consider the undirected and unweighted graph of Fig. 5a. Node $v_1$ has geodesic number 2 since the closest nodes with explicit beliefs are $v_2$ and $v_7$ two hops away. There are three highlighted shortest paths to those beliefs. The SBP standardized belief assignment is then $\hat{\mathbf{b}}'_{v_1} = \zeta(\hat{\mathbf{H}}_o^2(2\hat{\mathbf{e}}_{v_2} + \hat{\mathbf{e}}_{v_7}))$. Notice that the factor 2 for $\hat{\mathbf{e}}_{v_1}$ arises from the 2 shortest paths from $v_2$ to $v_1$.*

Given a graph with adjacency matrix $\mathbf{A}$ and a selection of explicit nodes. Then for any edge, one of two cases is true: (*i*) the edge connects two nodes with the same geodesic number, or (*ii*) the edge connects two nodes that have geodesic numbers of difference one. It follows that SBP has the same semantics as LinBP over a modified graph with some edges removed and the remaining edges becoming directed:

LEMMA 17 (MODIFIED ADJACENCY MATRIX). *Consider a graph with adjacency matrix $\mathbf{A}$ and a selection of explicit nodes. Remove all edges between nodes with same geodesic numbers. For the remaining edges, keep the direction from lower to higher geodesic number. Let $\mathbf{A}_*$ be the resulting modified adjacency matrix. Then: (1) the directed graph $\mathbf{A}_*$ has no directed cycles; and (2) SBP for $\mathbf{A}$ leads to the same final beliefs as LinBP over the transpose $\mathbf{A}_*^\intercal$.*

EXAMPLE 18 (SBP ADJACENCY MATRIX). *Let's consider again the undirected graph of Fig. 5b. Among the 4 entries for $v_1 - v_3$ and $v_1 - v_5$ in $\mathbf{A}$, the modified adjacency matrix contains only one entry for $v_3 \to v_1$, because $v_3, v_1, v_5$ have geodesic numbers 1, 2, 2, respectively. Thus the edge $v_1 - v_3$ only propagates information from $v_3$ to $v_1$, and the edge $v_1 - v_5$ propagates no information, as both end points have the same geodesic number.*

$$\mathbf{A} = \begin{bmatrix} 0 & 0 & 1 & 1 & 0 & 0 & 0 \\ 0 & 0 & 1 & 1 & 0 & 0 & 0 \\ 1 & 1 & 0 & 0 & 0 & 0 & 1 \\ 1 & 1 & 0 & 0 & 1 & 0 & 0 \\ 0 & 0 & 0 & 1 & 0 & 1 & 0 \\ 0 & 0 & 0 & 0 & 1 & 0 & 1 \\ 0 & 0 & 1 & 0 & 0 & 1 & 0 \end{bmatrix} \quad \mathbf{A}_* = \begin{bmatrix} 0 & 0 & 0 & 0 & 0 & 0 & 0 \\ 0 & 0 & 1 & 1 & 0 & 0 & 0 \\ 0 & 0 & 0 & 0 & 0 & 0 & 0 \\ 1 & 0 & 0 & 0 & 1 & 0 & 0 \\ 0 & 0 & 0 & 0 & 0 & 0 & 0 \\ 0 & 0 & 0 & 0 & 1 & 0 & 0 \\ 0 & 0 & 1 & 0 & 0 & 1 & 0 \end{bmatrix}$$

The following theorem gives the connection between LinBP and SBP and is the main result of this section.

THEOREM 19 (LIMIT OF LINBP). *For $\lim_{\epsilon_H \to 0^+}$, the standardized belief assignment for LinBP converges towards the standardized belief assignment for SBP.*

In other words, except for ties (!), the top belief assignment for LinBP and SBP are equal for sufficiently small $\epsilon_H$.

EXAMPLE 20 (DETAILED EXAMPLE). *Consider the unweighted and undirected torus graph shown in Fig. 5c, and assume explicit beliefs $\hat{\mathbf{e}}_{v_1} = [2, -1, -1]$, $\hat{\mathbf{e}}_{v_2} = [-1, 2, -1]$, $\hat{\mathbf{e}}_{v_3} = [-1, -1, 2]$, plus the coupling matrix from Fig. 1c. We get the unscaled residual matrix by centering all entries around $\frac{1}{3}$: $\hat{\mathbf{H}}_o = \begin{bmatrix} 0.6 & 0.3 & 0.1 \\ 0.3 & 0.0 & 0.7 \\ 0.1 & 0.7 & 0.2 \end{bmatrix} - \begin{bmatrix} \frac{1}{3} \end{bmatrix}_{3 \times 3}$. We focus on node $v_4$ and compare the standardized belief assignment $\hat{\mathbf{b}}'_{v_4}$ and the standard deviation $\sigma(\hat{\mathbf{b}}_{v_4})$ between BP, LinBP, LinBP*, and*

SBP for $\hat{\mathbf{H}} = \epsilon_H \hat{\mathbf{H}}_o$ and the limit of $\epsilon_H \to 0$. SBP predicts the standardized beliefs to result from the two shortest paths, $v_1 \to v_5 \to v_8 \to v_4$ and $v_3 \to v_7 \to v_8 \to v_4$, and thus $\hat{\mathbf{b}}'_{v_4} = \zeta(\hat{\mathbf{H}}_o^3 (\hat{\mathbf{e}}_{v_1} + \hat{\mathbf{e}}_{v_3})) \approx [-0.069, 1.258, -1.189]$. For the standard deviation, we get $\sigma(\hat{\mathbf{b}}_{v_4}) = \sigma(\hat{\mathbf{H}}^3 (\hat{\mathbf{e}}_{v_1} + \hat{\mathbf{e}}_{v_3})) = \epsilon_H^3 \sigma(\hat{\mathbf{H}}_o^3 (\hat{\mathbf{e}}_{v_1} + \hat{\mathbf{e}}_{v_3})) \approx \epsilon_H^3 \cdot 0.332$. According to Eq. 16, LinBP converges iff $\rho(\epsilon_H \hat{\mathbf{H}}_0 \otimes \mathbf{A} - \epsilon_H^2 \hat{\mathbf{H}}_0^2 \otimes \mathbf{D}) < 1$, from which we can calculate numerically $\epsilon_H \lesssim 0.488$. According to Eq. 17, LinBP* converges iff $\epsilon_H < \frac{1}{\rho(\hat{\mathbf{H}}_o)\rho(\mathbf{A})}$ and thus for $\epsilon_H \lesssim 0.658$, given $\rho(\mathbf{A}) \approx 2.414$ and $\rho(\hat{\mathbf{H}}_o) \approx 0.629$. Using the norm approximations instead of the spectral radii, we get $\epsilon_H \lesssim 0.360$ for LinBP, and $\epsilon_H \lesssim 0.455$ for LinBP* as sufficient (but not necessary) conditions for convergence. Figure 4c and Fig. 4d illustrate that our spectral radii criteria capture the convergence of LinBP and LinBP* exactly.

### 6.3 SBP in SQL

The SBP semantics may assign beliefs to a node that depend on an exponential number of paths (exponential in the geodesic number of a node). However, SBP actually allows a simple algorithm in SQL that propagates information across every edge at most once, which justifies our choice of name "single-pass". We achieve this in SQL by adding a table $G(\underline{v}, g)$ to the schema that stores the geodesic number $g$ for each node $v$. This table $G$, in turn, also supports efficient updates. In the following, we give two algorithms for (1) the initial assignments of beliefs and (2) addition of explicit beliefs. The Appendix also includes an algorithm for (3) addition of edges to the graph.

**(1) Initial belief assignment.** Algorithm 2 shows the initial calculation of all final beliefs: We start with nodes with explicit beliefs; i.e. geodesic number 0 (line 1). At each subsequent iteration (line 3), we then determine nodes with increasing geodesic number by following edges from previously inserted nodes (i.e. those with geodesic number smaller by 1), but ignoring nodes that have already been visited (i.e. those that are already in $G$) (line 4). Note that in a slight abuse of Datalog notation (and for the sake of conciseness), we allow negation on relational atoms with anonymous variables implying a nested not exist query.[17] The beliefs of the new nodes are then calculated by following all edges from nodes that have just been assigned their beliefs in the previous step (line 5). This is repeated for nodes with increasing geodesic numbers until the table $G$ remains unchanged (line 7).

PROPOSITION 21 (ALGORITHM 2). *Algorithm 2 terminates in finite number of iterations and returns a sound and complete enumeration of final beliefs according to SBP.*

**(2) Addition of explicit beliefs.** We assume the set of changed or additional explicit beliefs to be available in table $E_n(\underline{v}, c, b)$ and use tables $G_n(\underline{v}, g)$ and $B_n(\underline{v}, c, b)$ to store temporary information for nodes that get updated. We will further use an exclamation mark left of a Datalog query to imply that the respective data record is either inserted or an existing one updated. Algorithm 3 shows the SQL translation for batch updates of explicit beliefs: Line 1 and line 2 initialize tables $G_n$ and $B_n$ for all *new* explicit nodes. At

---

**Algorithm 2:** (SBP) Returns the final beliefs $B$ and geodesic numbers $G$ with SBP for a weighted network $A$ with explicit beliefs $E$, and coupling scores $H$.

**Input**: $A(s, t, w)$, $E(v, c, b)$, $H(c_1, c_2, h)$
**Output**: $B(v, c, b)$, $G(v, g)$

1 Initialize geodesic n. and beliefs for nodes with explicit beliefs:
 $G(v,' 0') \coloneq E(v, \_, \_)$
 $B(v, c, b) \coloneq E(v, c, b)$
2 $i \leftarrow 1$
3 **repeat**
4 | Find next nodes to calculate:
 $G(t, i) \coloneq G(s, i-1), A(s, t, \_), \neg G(t, \_)$
5 | Calculate beliefs for new nodes:
 $B(t, c_2, sum(w \cdot b \cdot h)) \coloneq G(t, i), A(s, t, w), B(s, c_1, b),$
 $G(s, i-1), H(c_1, c_2, h)$
6 | $i \leftarrow i + 1$
7 **until** *no more inserts into $G$*
8 **return** $B$ and $G$

---

**Algorithm 3:** ($\Delta$SBP:newExplicitBeliefs) Updates $B$ and $G$, given new explicit beliefs $E_n$ and weighted network $A$.

**Input**: $E_n(v, c, b)$, $A(s, t, w)$
**Output**: Updated $B(v, c, b)$ and $G(v, g)$

1 Initialize geodesic numbers for new nodes with explicit beliefs:
 $G_n(v,' 0') \coloneq E_n(v, \_, \_)$
 $! G(v,' 0') \coloneq G_n(v, \_)$
2 Initialize beliefs for new nodes:
 $B_n(v, c, b) \coloneq E_n(v, c, b)$
 $! B(v, c, b) \coloneq B_n(v, c, b)$
3 $i \leftarrow 1$
4 **repeat**
5 | Find next nodes to update:
 $G_n(t, i) \coloneq G_n(s, i-1), A(s, t, \_), \neg(G(t, g_t), g_t < i)$
 $! G(v, i) \coloneq G_n(v, i)$
6 | Calculate new beliefs for these nodes:
 $B_n(t, c_2, sum(w \cdot b \cdot h)) \coloneq G_n(t, i), A(s, t, w), B(s, c_1, b),$
 $G(s, i-1), H(c_1, c_2, h)$
 $! B(v, c, b) \coloneq B_n(v, c, b)$
7 | $i \leftarrow i + 1$
8 **until** *no more inserts into $G_n$*
9 **return** $B$ and $G$

---

each subsequent iteration $i$ (line 4), we then determine all nodes $t$ that need to be updated with *new* geodesic number $g_t = i$ by following edges from previously updated nodes $s$ with geodesic number $g_s = i - 1$ and ignoring those that already have a smaller geodesic number $g_t < i$. (line 5).[18] For these nodes $t$, the updated beliefs are then calculated by only following edges that start at nodes $s$ with geodesic number $g_s = i - 1$, independent of whether those were updated or not (line 6). The algorithm terminates when there are no more inserts in table $G_n$ (line 8).

PROPOSITION 22 (ALGORITHM 3). *Algorithm 3 terminates in finite number of iterations and returns a sound and complete enumeration of updated beliefs.*

### 7. EXPERIMENTS

---

[17]The common syntactic safety restriction is that all variables need to appear in a positive relational atom of the body. In practice, we use a left outer join and an "is null" condition.

[18]Note that edges $s \to t$ with $g_s \geq g_t$ cannot contain a geodesic path in that direction and are thus ignored. Also note that, again for the sake of conciseness, we write $\neg(G(t, g), g < i)$ to indicate that nodes $t$ with $g_t < i$ are not updated. In SQL, we used an except clause.

| | Graph characteristics | | | Explicit b. | |
|---|---|---|---|---|---|
| # | Nodes $n$ | Edges $e$ | $e/n$ | 5% | 1‰ |
| 1 | 243 | 1 024 | 4.2 | 12 | 1 |
| 2 | 729 | 4 096 | 5.6 | 36 | 1 |
| 3 | 2 187 | 16 384 | 7.6 | 110 | 3 |
| 4 | 6 561 | 65 536 | 10.0 | 328 | 7 |
| 5 | 19 683 | 262 144 | 13.3 | 984 | 20 |
| 6 | 59 049 | 1 048 576 | 17.8 | 2 952 | 60 |
| 7 | 177 147 | 4 194 304 | 23.7 | 8 857 | 178 |
| 8 | 531 441 | 16 777 216 | 31.6 | 26 572 | 532 |
| 9 | 1 594 323 | 67 108 864 | 42.6 | 79 716 | 1 595 |

(a) Number of nodes, edges, explicit beliefs

| | 1 | 2 | 3 |
|---|---|---|---|
| 1 | 10 | -4 | -6 |
| 2 | -4 | 7 | -3 |
| 3 | -6 | -3 | 9 |

(b) Unscaled residual coupling m. $\hat{\mathbf{H}}_o$

Figure 6: Synthetic data used for our experiments.

In this section, we experimentally verify how well our new methods LinBP and SBP scale, and how close the top belief classification of both methods matches that of standard BP.

**Experimental setup.** We implemented main memory-based versions of BP and LinBP in JAVA, and disk-bound versions of LinBP and SBP in SQL. The JAVA implementation uses optimized libraries for sparse matrix operations [37]. When timing our memory-based algorithms, we focus on the running times for computations only and ignore the time for loading data and initializing matrices. For the SQL implementation, we report the times from start to finish on PostgreSQL 9.2 [39]. We are mainly interested in relative performance within a platform (LinBP vs. BP in JAVA, and SBP vs. LinBP in SQL) and scalability with graph sizes. Both implementations run on a 2.5 Ghz Intel Core i5 with 16G of main memory and a 1TB SSD hard drive. To allow comparability across implementations, we limit evaluation to one processor. For timing results, we run BP and LinBP for 5 iterations, and SBP until termination.

**Synthetic Data.** We assume a scenario with $k = 3$ classes and the matrix $\hat{\mathbf{H}}_o$ from Fig. 6b as the unscaled coupling matrix. We study the convergence of our algorithms by scaling $\hat{\mathbf{H}}_o$ with a varying parameter $\epsilon_H$. We created 9 "Kronecker graphs" of varying sizes (see Fig. 6a) which are known to share many properties with real world graphs [28].[19] To generate initial class labels (explicit beliefs), we pick 5% of the nodes in each graph and assign to them two random numbers from $\{-0.1, -0.09, \ldots, 0.09, 0.1\}$ as centered beliefs for two classes (the belief in the third class is then their negative sum due to centering). For timing of incremental updates for SBP (denoted as $\Delta$SBP), we created similar updates for 2% of the nodes with explicit beliefs (corresponding to 1‰ = 0.1% of all nodes in a graph).

**Measuring classification quality.** We take the top beliefs returned by BP as "ground truth" (GT) and are interested in how close the classifications returned by LinBP and SBP come for varying scaling of $\hat{\mathbf{H}}_o$.[20] We measure quality of our methods with *precision* and *recall* as follows: Given a set of top beliefs $B_{\text{GT}}$ for a GT labeling method and a set of top beliefs $B_O$ of another method (O), let $B_\cap$ be the set of shared beliefs: $B_\cap = B_{\text{GT}} \cap B_O$. Then, recall $r$ measures the portion of GT beliefs that are returned by O: $r = |B_\cap|/|B_{\text{GT}}|$, and precision $p$ measures the portion of "correct" beliefs among $B_O$: $p = |B_\cap|/|B_O|$. Notice that this method naturally handles ties. For example, assume that the GT assigns classes $c_1, c_2, c_3$ as top beliefs to 3 nodes $v_1, v_2, v_3$, respectively: $\{v_1 \to c_1, v_2 \to c_2, v_3 \to c_3\}$, whereas the comparison method assigns 4 beliefs: $\{v_1 \to \{c_1, c_2\}, v_2 \to c_2, v_3 \to c_2\}$. Then $r = 2/3$ and $p = 2/4$.

QUESTION 1. *Timing: How fast and scalable are LinBP and SBP as compared to BP in both implementations?*

*Result 1.* The main memory implementation of LinBP is up to 600 times faster than BP, and the SQL implementation of SBP is more than 10 times faster than LinBP.

Figure 7a and Fig. 7b show our timing experiments in both JAVA and SQL, respectively. Figure 7c shows the times for the 5 largest graphs. Notice that all implementations except BP show approximate linear scaling behavior in the number of edges (as reference, both Fig. 7a and Fig. 7b show a dashed grey line that represents an exact linear scalability of 100 000 edges per second). The main-memory implementation of LinBP is 600 times faster than that of BP for the largest graph. We see at least two reasons for these speed-ups: (*i*) the LinBP update equations calculate beliefs as function of beliefs. In contrast, the BP update equations calculate, for each node, outgoing messages as function of incoming messages; (*ii*) our matrix formulation of LinBP enables us to use well-optimized JAVA libraries for matrix operations. These optimized operations lead to a highly efficient algorithm. SBP is 10 times faster than LinBP in SQL (we look at this closer in the next question). Not surprisingly, the main-memory JAVA implementation of LinBP is much faster than the disk-bound LinBP implementation in SQL. It is worth mentioning that even though our SQL implementation did not exploit special libraries and is the disk-bound, our SBP implementation in SQL is still faster than the BP implementation in JAVA (!).

QUESTION 2. *Timing: What can the speed-up of SBP over LinBP be mostly attributed to?*

*Result 2.* SBP needs fewer iterations to converge and requires fewer calculations for each iteration, on average.

Figure 7d shows the time required by our JAVA implementation for both LinBP and SBP within each iteration on graph #7. SBP visits different edges in each iteration, and thus needs a different amount of time for each iteration, whereas LinBP revisits every edge in every iteration again. The fact that SBP needs more time for the 2nd iteration than LinBP, although fewer edges are visited, is a consequence of the overhead for maintaining the indexing structure required to decide on which edges to visit next.

QUESTION 3. *Timing: When is it faster to update a graph incrementally than to recalculate from scratch with SBP?*

*Result 3.* In our experiments, it was faster to update SBP when less than ≈ 50% of the final explicit beliefs are new.

Figure 7e shows the results for SQL on graph #5. We fix 10% of the nodes with explicit beliefs after the update. Among these nodes, we vary a certain fraction as *new* beliefs. For example, 20% on the horizontal axis implies that

---

[19] Notice that we count the number of entries in **A** as the number of edges; thus, each edge is counted twice ($s-t$ equals $s \to t$ plus $t \to s$).

[20] Our experimental approach is justified since BP has previously been shown to work well in real-life classification scenarios. Our goal in this paper is not to justify BP for such inference, but rather to replace BP with a faster and simpler semantics that gives similar classifications.

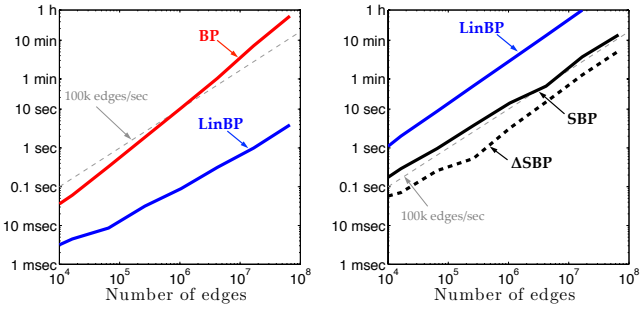
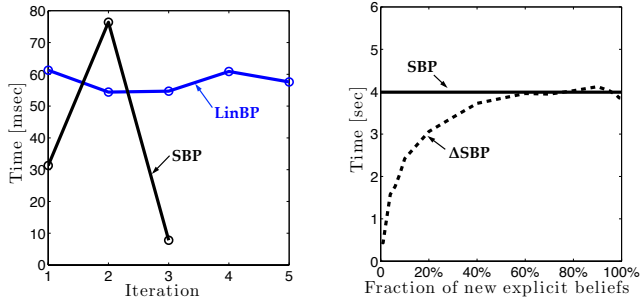
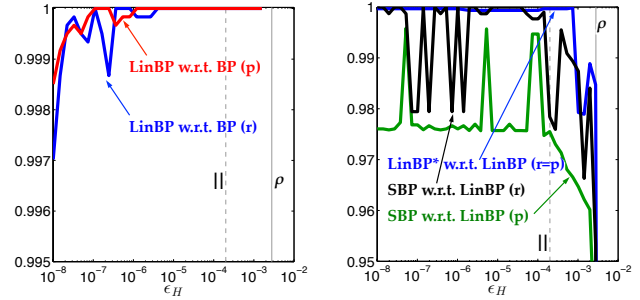

(a) Scalability JAVA  (b) Scalability SQL  (c) Timing results of all methods in SQL/JAVA on 5 largest graphs

(d) SBP/LinBP in JAVA on #7  (e) ΔSBP/SBP in SQL on #5  (f) Recall & precision on #5  (g) Recall & precision on #5

Figure 7: (a)-(c): Scalability of methods in Java and SQL: dashed gray lines represent linear scalability. (d): ΔSBP vs. SBP for various fractions of updates assuming 10% explicit beliefs. (e): Timing LinBP and SBP per iteration. (f),(g): Quality of LinBP w.r.t. BP, and SBP w.r.t. LinBP: the vertical gray lines mark $\epsilon_H = 0.0002$, i.e. the sufficiency convergence criterium from Lemma 9.

we had 80% of the explicit nodes (= 8% of all nodes) known before the update, and are now adding 20% of the explicit nodes (= 2% of all nodes) with the incremental SBP Algorithm 3 ("ΔSBP"). For the alternative Algorithm 2 ("SBP"), we recalculate the final beliefs from scratch (therefore, shown with a constant horizontal line). In addition, Fig. 7c shows timing for updating 1‰ of the nodes in a graph that previously had 5% nodes with explicit beliefs: The relative speed-up is around 2.5 for the larger graphs.

QUESTION 4. *Quality: How do the top belief assignments of LinBP, LinBP* and SBP compare to that of BP?*

*Result 4.* BP, LinBP, LinBP*, and SBP give almost identical top belief assignments with $\epsilon_H$ given by Lemma 9.

Figure 7f shows recall ($r$) and precision ($p$) of LinBP with BP as GT ("LinBP with regard to BP") on graph #5 (similar results hold for all other graphs). The vertical gray lines show $\epsilon_H = 0.0002$ and $\epsilon_H = 0.0028$, which result from our sufficient (Lemma 9) and exact (Lemma 8) convergence criteria of LinBP, respectively. The graphs stop earlier than $\epsilon_H = 0.0028$ as BP stops converging earlier. We see that LinBP matches the top belief assignment of BP exactly in the upper range of guaranteed convergence; for smaller $\epsilon_H$, errors result from roundoff errors due to limited precision of floating-point computations. We thus recommend choosing $\epsilon_H$ according to Lemma 8. Overall accuracy (harmonic mean of precision and recall) is still > 99.9% across all $\epsilon_H$.

Figure 7g shows that the results of LinBP and LinBP* are almost identical as long as $\epsilon_H$ is small enough for the algorithms to converge (both LinBP and LinBP* always return unique top belief assignments; thus, $r$ and $p$ are identical and we only need to show one graph for both). The vertical drops in $r$ and $p$ on the right correspond to choices of $\epsilon_H$ for which LinBP stops converging.

Figure 7g also validates that SBP closely matches LinBP (and thus BP). The averaged recall of SBP w.r.t. LinBP for $10^{-9} < \epsilon_H < 0.0002$ is 0.995 and the averaged precision 0.978. Thus overall accuracy is > 98.6% across all $\epsilon_H$. The visible oscillations and the observation that SBP's precision values are generally lower than its recall values are mainly due to tied top beliefs for SBP. In such a case, SBP returns two top beliefs, while LinBP returns only one. For example, we observed the following final beliefs which lead to a drop in precision (due to SBP's tie):
- LinBP: $[1.0000000014, 1.0000000002, -2.0000000016] \cdot 10^{-2}$
- SBP: $[1, 1, -2] \cdot 10^{-2}$

The following more rare scenario is due to numerical rounding errors and led to a drop in both precision and recall (LinBP and SBP return two different top beliefs):
- LinBP: $[7.60009, 7.60047, -15.20056] \cdot 10^{-11}$
- SBP: $[7.6, 7.59999999999999, -15.2] \cdot 10^{-11}$

Minimizing the possibility of ties by choosing initial explicit beliefs with additional digits (e.g., 0.0503 instead of 0.05) removed these oscillations.

In summary, SBP and LinBP match the classification of BP very well. Misclassifications are mostly due to closely tied top beliefs, in which case returning both tied beliefs (as done by SBP) would arguably be *the preferable alternative.*

## 8. RELATED WORK

The two main philosophies for transductive inference are *logical* approaches and *connectionist* approaches (see Fig. 8).

**Logical approaches** determine the solution based on hard rules, and are most common in the database literature. Examples are trust mappings, preference-based up-dates, stable model semantics, but also tuple-generating dependencies, inconsistency-resolution, database repairs, community databases. Example applications are peer-data management and collaborative data sharing systems that have to deal with conflicting data and lack of consensus about which data is correct during integration, update exchange, and that have adopted some form of conflict handling or trust mappings in order to facilitate data sharing among users [3, 11, 13, 15, 16, 22, 23, 43]. Commonly, those inconsistencies are expressed with key violations [10] and resolved at query time through database repairs [1].

| Databases | Machine Learning |
|---|---|
| Inconsistency resolution | Semi-supervised learning |
| Logic-based approaches | Connectionist approaches |
| extensional database | prior beliefs |
| intensional database | posterior beliefs |

Figure 8: Comparing common formulations of transductive inference in the database vs. machine learning communities.

**Connectionist approaches** determine the solution based on soft rules. The related work comprises guilt-by-association approaches, which use limited prior knowledge and network effects in order to derive new knowledge. The main alternatives are semi-supervised learning (SSL), random walks with restarts (RWR), and label or belief propagation (BP). SSL methods can be divided into low-density separation methods, graph-based methods, methods for changing the representation, and co-training methods (see [29, 47] for overviews). A multi-class approach has been introduced in [20]. RWR methods are used to compute mainly node relevance; e.g., original and personalized PageRank [4, 17], lazy random walks [31], and fast approximations [35, 44].

**Belief Propagation** (or min-sum or product-sum algorithm) is an iterative message-passing algorithm that is a very expressive formalism for assigning classes to unlabeled nodes and has been used successfully in multiple settings for solving inference problems, such as error-correcting codes [27] or stereo imaging in computer vision [9], fraud detection [30, 36], malware detection[5], graph similarity [2, 26], structure identification [24], and pattern mining and anomaly detection [21]. BP solves the inference problem approximately; it is known that when the factor graph has a tree structure, it reaches a stationary point (convergence to the true marginals) after a finite number of iterations. Although in loopy factor graphs, convergence to the correct marginals is not guaranteed, the true marginals may still be achieved in *locally* tree-like structures. As a consequence, approaches in the database community that rely on BP-type of inference also commonly lack convergence guarantees [42].

Convergence of BP in loopy graphs has been studied before [8, 19, 32]. To the best of our knowledge, all existing bounds for BP give only sufficient convergence criteria. In contrast, our work presents a stronger result by providing sufficient and necessary conditions for the convergence of LinBP, which is itself an approximation for BP.

There exist various works that speed up BP by: (*i*) exploiting the graph structure [6, 36], (*ii*) changing the order of message propagation [8, 14, 32], or (*iii*) using the MapReduce framework [21]. Here, we derive a linearized formulation of standard BP. This is a multivariate (polytomous) generalization of the linearized belief propagation algorithm FABP [25] from binary to multiple labels for classification. In addition, we provide translations into SQL and a new, faster semantics that captures the underlying intuition and provides efficient incremental updates.

**Incremental maintenance.** While the nearest-labeled-neighbor-semantics allows efficient incremental updates for SBP (cf. Lemma 17), incrementally updating the result of LinBP is more challenging since it involves general matrix computations. For such scenarios, combining our work with approaches like the one from [34] is left for future work.

## 9. CONCLUSIONS

This paper showed that the widely used multi-class belief propagation algorithm can be approximated by a linear system that replaces multiplication with addition. This allows us to give a fast and compact matrix formulation and a compact implementation in standard SQL. The linear system also allows a closed-form solution with the help of the inverse of an appropriate matrix. We can thus explain *exactly* when the system will converge, and what the limit value is as the neighbor-to-neighbor influence tends to zero. For the latter case, we show that the scores depend only on the "nearest labeled neighbor," which leads to an even faster algorithm that also supports incremental updates.

**Acknowledgements.** We would like to thank Garry Miller for pointing us to Roth's column lemma and the anonymous reviewers for their careful reading and detailed feedback.

# APPENDIX

## A. NOMENCLATURE

| | |
|---|---|
| $n$ | number of nodes |
| $s, t, u$ | indices used for nodes |
| $N(s)$ | list of neighbours for node s |
| $k$ | number of classes |
| $i, j, g$ | indices used for classes |
| $\mathbf{e}_s$ | $k$-dimensional prior (explicit) belief vector at node $s$ |
| $\mathbf{b}_s$ | $k$-dim. posterior (implicit, final) belief vector at node $s$ |
| $\mathbf{m}_{st}$ | $k$-dim. message vector from node $s$ to node $t$ |
| $\mathbf{A}$ | $n \times n$ weighted symmetric adjacency matrix with $A(s,t)$ being the weight for edge $s \to t$ |
| $\mathbf{E}, \mathbf{B}$ | $n \times k$ explicit or implicit belief matrix with $E(s,i)$ indicating the strength of belief in class $i$ by node $s$ |
| $\mathbf{H}$ | $k \times k$ coupling matrix with $H(j,i)$ indicating the influence of class $j$ of a sender on class $i$ of the recipient |
| $\hat{\mathbf{H}}, \hat{\mathbf{E}}, \hat{\mathbf{B}}$ | residual matrices centered around $\frac{1}{k}$ |
| $\hat{\mathbf{H}}_o$ | unscaled, original coupling matrices $\hat{\mathbf{H}} = \epsilon_H \hat{\mathbf{H}}_o$ |
| $\epsilon_H$ | scaling factor |
| $\mathbf{I}_k$ | $k$-dimensional identity matrix |
| $\mathbf{X}^\intercal$ | transpose of matrix $\mathbf{X}$ |
| $\text{vec}(\mathbf{X})$ | vectorization of matrix $\mathbf{X}$ |
| $\mathbf{X} \otimes \mathbf{Y}$ | Kronecker product between matrices $\mathbf{X}$ and $\mathbf{Y}$ |
| $\frac{1}{Z}$ | normalizer |

## B. PROOFS

### B.1 Proofs for Sect. 4: Derivation of LinBP

**Lemma 5: Centered BP.**

PROOF LEMMA 5. Equation 8: Substituting the expansions into the belief updates Eq. 1 leads to

$$\frac{1}{k} + \hat{b}_s(i) \leftarrow \frac{1}{Z_s} \cdot \left(\frac{1}{k} + \hat{e}_s(i)\right) \cdot \prod_{u \in N(s)} \left(1 + \hat{m}_{us}(i)\right)$$

$$\ln\left(1 + k\hat{b}_s(i)\right) \leftarrow -\ln Z_s + \ln\left(1 + k\hat{e}_s(i)\right) + \sum_{u \in N(s)} \ln\left(1 + \hat{m}_{us}(i)\right)$$

$$k\hat{b}_s(i) \leftarrow -\ln Z_s + k\hat{e}_s(i) + \sum_{u \in N(s)} \hat{m}_{us}(i) \quad (22)$$

For the last step, we use the approximation $\ln(1 + \epsilon) \approx \epsilon$ for small $\epsilon$. Summing both sides over $i$ gives us:

$$k \underbrace{\sum_i \hat{b}_s(i)}_{=0} \leftarrow -k \ln Z_s + k \underbrace{\sum_i \hat{e}_s(i)}_{=0} + \underbrace{\sum_i \sum_{u \in N(s)} \hat{m}_{us}(i)}_{=0}$$

Hence, we see that $\ln Z_s$ needs to be 0, and therefore our normalizer is actually a normalization constant and independent for all nodes $Z_s = 1$. Plugging $Z_s = 1$ back into Eq. 22 leads to Eq. 8:

$$\hat{b}_s(i) \leftarrow \hat{e}_s(i) + \frac{1}{k} \sum_{u \in N(s)} \hat{m}_{us}(i)$$

Equation 9: We first write Eq. 3 as:

$$m_{st}(i) \leftarrow \frac{1}{Z_{st}} \sum_j H(j, i) \, e_s(j) \prod_{u \in N(s) \setminus t} m_{us}(j) \quad (23)$$

$$\leftarrow \frac{Z_s}{Z_{st}} \sum_j H(j, i) \frac{\frac{1}{Z_s} e_s(j) \prod_{u \in N(s)} m_{us}(j)}{m_{ts}(j)}$$

$$\leftarrow \frac{Z_s}{Z_{st}} \sum_j H(j, i) \frac{b_s(j)}{m_{ts}(j)} \quad (24)$$

Then, using, $Z_s = 1$ and the expansions together with the approximation $\frac{\frac{1}{k} + \epsilon_1}{1 + \epsilon_2} \approx \frac{1}{k} + \epsilon_1 - \frac{1}{k}\epsilon_2$ for small $\epsilon_1, \epsilon_2$, we get:

$$1 + \hat{m}_{st}(i) \leftarrow \frac{1}{Z_{st}} \sum_j \left(\frac{1}{k} + \hat{H}(j, i)\right) \frac{\frac{1}{k} + \hat{b}_s(j)}{1 + \hat{m}_{ts}(j)}$$

$$\leftarrow \frac{1}{Z_{st}} \sum_j \left(\frac{1}{k} + \hat{H}(j, i)\right)\left(\frac{1}{k} + \hat{b}_s(j) - \frac{1}{k}\hat{m}_{ts}(j)\right)$$

$$\leftarrow \frac{1}{Z_{st}} \left(\frac{1}{k} + \frac{1}{k}\underbrace{\sum_j \hat{H}(j,i)}_{=0} + \frac{1}{k}\underbrace{\sum_j \hat{b}_s(j)}_{=0} + \sum_j \hat{H}(j,i)\hat{b}_s(j)\right.$$

$$\left. - \frac{1}{k^2}\underbrace{\sum_j \hat{m}_{ts}(j)}_{=0} - \frac{1}{k}\sum_j \hat{H}(j,i)\hat{m}_{ts}(j)\right) \quad (25)$$

We can then determine the normalization factor $Z_{st}$ to be a constant as well ($Z_{st} = k^{-1}$) by summing both sides of Eq. 25 over $i$ and observing that $\sum_j \hat{b}_s(j) \sum_i \hat{H}(j, i) = 0$, since $\sum_i \hat{H}(j, i) = 0$:

$$k + \underbrace{\sum_i \hat{m}_{st}(i)}_{=0} \leftarrow \frac{1}{Z_{st}} \cdot \Big(1 + \underbrace{\sum_i \sum_j \hat{H}(j,i)\hat{b}_s(j)}_{=0}$$

$$- \underbrace{\sum_i \sum_j \hat{H}(j,i)\hat{m}_{ts}(j)}_{=0}\Big)$$

We get Eq. 9 from Eq. 25 and $\frac{1}{Z_{st}} = k$.

$$\hat{m}_{st}(i) \leftarrow k \sum_j \hat{H}(j,i)\hat{b}_s(j) - \sum_j \hat{H}(j,i)\hat{m}_{ts}(j)$$

□

**Lemma 6: Steady state messages.**

PROOF LEMMA 6. Using Eq. 9 and plugging for $\hat{m}_{ts}(j)$ back into the equation for $\hat{m}_{st}(j)$, we get:

$$\hat{m}_{st}(i) \leftarrow k\sum_j \hat{H}(j,i)\hat{b}_s(j) - \sum_j \hat{H}(j,i) \cdot$$

$$\left(k\sum_g \hat{H}(g,j)\hat{b}_t(g) - \sum_g \hat{H}(g,j)\hat{m}_{st}(g)\right)$$

Now, for the case of convergence, both $\hat{m}_{st}(g)$ on the left and right side of the equation need to be equivalent. We can, therefore, group related terms together and replace the update symbol with equality:

$$\hat{m}_{st}(i) - \sum_j \hat{H}(j,i) \sum_g \hat{H}(g,j)\hat{m}_{st}(g)$$

$$= k\sum_j \hat{H}(j,i)\hat{b}_s(j) - k\sum_j \hat{H}(j,i)\sum_g \hat{H}(g,j)\hat{b}_t(g) \quad (26)$$

This equation can then be written in matrix notation as:

$$(\mathbf{I}_k - \hat{\mathbf{H}}^2)\hat{\mathbf{m}}_{st} = k\hat{\mathbf{H}}\hat{\mathbf{b}}_s - k\hat{\mathbf{H}}^2\hat{\mathbf{b}}_t \quad (27)$$

which leads to Eq. 10, given that all entries of $\hat{\mathbf{H}} \ll \frac{1}{k}$ and thus the inverse of $(\mathbf{I}_k - \hat{\mathbf{H}}^2)$ always exists. □

**Theorem 4: Linearized BP (LinBP).**

PROOF THEOREM 4. For steady-state, we can write Eq. 8 in vector form as:

$$\hat{\mathbf{b}}_s = \hat{\mathbf{e}}_s + \frac{1}{k}\sum_{u \in N(s)} \hat{\mathbf{m}}_{us}$$

and by substituting $\hat{\mathbf{H}}_*$ for $(\mathbf{I}_k - \hat{\mathbf{H}}^2)^{-1}\hat{\mathbf{H}}$, we write Eq. 10 as

$$\hat{\mathbf{m}}_{us} = k\hat{\mathbf{H}}_*(\hat{\mathbf{b}}_u - \hat{\mathbf{H}}\hat{\mathbf{b}}_s)$$

Combining the last two equations, we get

$$\hat{\mathbf{b}}_s = \hat{\mathbf{e}}_s + \hat{\mathbf{H}}_* \sum_{u \in N(s)} \hat{\mathbf{b}}_u - d_s \hat{\mathbf{H}}_* \hat{\mathbf{H}} \hat{\mathbf{b}}_s \quad (28)$$

where $d_s$ is the degree or number of *bi-directional* neighbors for node $s$, i.e. neighbors that are connected to $s$ with edges in both directions (see Sect. 5.2 for a discussion of the implication). By using $\hat{\mathbf{B}}$ and $\hat{\mathbf{E}}$ as $n \times k$ matrices of final and initial beliefs, $\mathbf{D}$ as the diagonal degree matrix, and $\mathbf{A}$ as the adjacency matrix, Eq. 28 can be written in matrix form

$$\hat{\mathbf{B}} = \hat{\mathbf{E}} + \mathbf{A}\hat{\mathbf{B}}\hat{\mathbf{H}}_* - \mathbf{D}\hat{\mathbf{B}}\hat{\mathbf{H}}\hat{\mathbf{H}}_* \quad (29)$$

By approximating $(\mathbf{I}_k - \hat{\mathbf{H}}^2) \approx \mathbf{I}_k$ (recall that all entries of $\hat{\mathbf{H}} << \frac{1}{k}$), and thus $\hat{\mathbf{H}}_* \approx \hat{\mathbf{H}}$, we can simplify to

$$\hat{\mathbf{B}} = \hat{\mathbf{E}} + \mathbf{A}\hat{\mathbf{B}}\hat{\mathbf{H}} - \mathbf{D}\hat{\mathbf{B}}\hat{\mathbf{H}}^2$$

And by further ignoring the second term with residual terms of third order, we can further simplify to get Eq. 5. □

**Proposition 7: LinBP in closed-form.**

PROOF PROPOSITION 7. *Roth's column lemma* [18] states that

$$\text{vec}(\mathbf{XYZ}) = (\mathbf{Z}^\intercal \otimes \mathbf{X})\text{vec}(\mathbf{Y})$$

With $\hat{\mathbf{H}}^\intercal = \hat{\mathbf{H}}$, we can thus write Eq. 4 as

$$\text{vec}(\hat{\mathbf{B}}) = \text{vec}(\hat{\mathbf{E}}) + (\hat{\mathbf{H}} \otimes \mathbf{A})\text{vec}(\hat{\mathbf{B}}) - (\hat{\mathbf{H}}^2 \otimes \mathbf{D})\text{vec}(\hat{\mathbf{B}})$$
$$= \text{vec}(\hat{\mathbf{E}}) + (\hat{\mathbf{H}} \otimes \mathbf{A} - \hat{\mathbf{H}}^2 \otimes \mathbf{D})\text{vec}(\hat{\mathbf{B}}) \quad (30)$$

which can be solved for $\text{vec}(\hat{\mathbf{B}})$ to get Eq. 11. □

## B.2 Proofs for Sect. 5: Benefits of LinBP

**Lemma 8: Convergence.**

PROOF LEMMA 8. From the Jacobi method for solving linear systems [40], we know that the update equation Eq. 14 converges if and only if the spectral radius of the matrix is smaller than 1. Thus, the criterion (Eq. 16) for LinBP (Eq. 11 follows immediately.

For Eq. 12, we have $\mathbf{M} = \hat{\mathbf{H}} \otimes \mathbf{A}$ and therefore $\rho(\hat{\mathbf{H}} \otimes \mathbf{A}) = \rho(\hat{\mathbf{H}})\rho(\mathbf{A}) < 1$, which holds if and only if $\rho(\hat{\mathbf{H}}) < \frac{1}{\rho(\mathbf{A})}$. □

Notice that Eq. 16 is an implicit criterion for $\hat{\mathbf{H}}$. We can given an alternative explicit sufficient (but not necessary) criterion as follows: we have $\mathbf{M} = \hat{\mathbf{H}} \otimes \mathbf{A} - \hat{\mathbf{H}}^2 \otimes \mathbf{D}$ and therefore $\rho(\hat{\mathbf{H}} \otimes \mathbf{A} - \hat{\mathbf{H}}^2 \otimes \mathbf{D}) \leq \rho(\hat{\mathbf{H}} \otimes \mathbf{A}) + \rho(\hat{\mathbf{H}}^2 \otimes \mathbf{D}) = \rho(\hat{\mathbf{H}})\rho(\mathbf{A}) + \rho(\hat{\mathbf{H}}^2)\rho(\mathbf{D}) \leq \rho(\hat{\mathbf{H}})\rho(\mathbf{A}) + \rho(\hat{\mathbf{H}})^2\rho(\mathbf{D}) < 1$, which holds if $\rho(\hat{\mathbf{H}}) \leq \frac{\sqrt{\rho(\mathbf{A})^2 + 4\rho(\mathbf{D})} - \rho(\mathbf{A})}{2\rho(\mathbf{D})}$.

**Lemma 9: Sufficient convergence criteria.**

PROOF LEMMA 9. Since $\rho(\mathbf{X}) \leq ||\mathbf{X}||$, it is sufficient to show that $||\mathbf{X}|| < 1$. For Eq. 11, we have $\rho(\hat{\mathbf{H}} \otimes \mathbf{A}) = \rho(\hat{\mathbf{H}})\rho(\mathbf{A}) \leq ||\hat{\mathbf{H}}||_i ||\mathbf{A}||_j < 1$, which holds if $||\hat{\mathbf{H}}||_i < \frac{1}{||\mathbf{A}||_j}$. Note that we can use different norms $||\cdot||_i$ and $||\cdot||_j$, and we get the best bounds for minimizing each norm individually.

For Eq. 12, we have $\rho(\hat{\mathbf{H}} \otimes \mathbf{A} - \hat{\mathbf{H}}^2 \otimes \mathbf{D}) \leq \rho(\hat{\mathbf{H}})\rho(\mathbf{A}) + \rho(\hat{\mathbf{H}})^2\rho(\mathbf{D}) \leq ||\hat{\mathbf{H}}||_i ||\mathbf{A}||_j + ||\hat{\mathbf{H}}||_i^2 ||\mathbf{D}||_k < 1$, which holds if $||\hat{\mathbf{H}}||_i \leq \frac{\sqrt{||\mathbf{A}||_j^2 + 4||\mathbf{D}||_k} - ||\mathbf{A}||_j}{2||\mathbf{D}||_k}$. Just as before, we can use different norms $||\cdot||_i, ||\cdot||_j$, and $||\cdot||_k$, and we get the best bounds for minimizing each norm individually. □

We also give an additional, simpler yet less tight sufficient condition for convergence of LinBP.

LEMMA 23 (ALTERNATIVE NORM CRITERION). *Let $||\cdot||$ stand for the induced 1-norm or induced $\infty$-norm of the enclosed matrix. Then LinBP converges if $||\hat{\mathbf{H}}|| < \frac{1}{2||\mathbf{A}||}$.*

PROOF LEMMA 23. For the induced 1-norm or $\infty$-norm (which are the maximum absolute column or row sum of a matrix, respectively), we know from the definition of $\mathbf{D}$, that $||\mathbf{D}|| \leq ||\mathbf{A}||$. With $||\hat{\mathbf{H}}||^2 < ||\hat{\mathbf{H}}|| < 1$, we thus have $||\hat{\mathbf{H}}|| ||\mathbf{A}|| + ||\hat{\mathbf{H}}||^2 ||\mathbf{D}|| < 2||\hat{\mathbf{H}}|| ||\mathbf{A}|| < 1$, from which $||\hat{\mathbf{H}}|| < \frac{1}{2||\mathbf{A}||}$. □

## B.3 Proofs for Sect. 6: SBP

**Lemma 17: Modified adjacency matrix.**

PROOF LEMMA 17. The fact that any remaining edge can go from a node with geodesic number $g$ to another node with geodesic number $g + 1$ follows immediately from a proof by contradiction. Assume there is an edge between $v_1$ and $v_2$ with $g(v_2) > g(v_1) + 1$. Then $v_2$ could reduce its geodesic number by following the edge between $v_1$ and $v_2$ which is a contradiction of the definition of the geodesic number.

Notice that $\mathbf{A}_*^\intercal$ is the actual matrix that is used by SBP instead of $\mathbf{A}$. It follows that the remaining graph cannot have any cycles as any path connects nodes with increasing geodesic numbers. A path can therefore never revisit a previous node.

The equivalence of SBP over $\mathbf{A}$ with LinBP over $\mathbf{A}_*^\intercal$ then follows from the fact that $\mathbf{A}_*^\intercal$ can only propagate information between two nodes $v_1$ and $v_2$ if $g(v_2) = g(v_1) + 1$. The necessary use of the transpose follow from the definition of a matrix multiplication and our convention that $A_*(s,t) \neq 1$ if the edge $s \to t$ exists (see Fig. 3). □

**Theorem 19: Convergence of LinBP towards SBP.**

PROOF THEOREM 19. Given an unscaled coupling matrix $\hat{\mathbf{H}}_o$, Eq. 28 for LinBP can be written as

$$\hat{\mathbf{b}}_s = \hat{\mathbf{e}}_s + \epsilon_H \hat{\mathbf{H}}_o \sum_{t \in N(s)} w_{ts}\hat{\mathbf{b}}_t$$

where $w_{us} = A(u,s)$ is the weight of the edge $u \to s$. Let $k$ be the geodesic number of node $s$.

If $k = 0$, i.e. node $s$ has explicit beliefs and thus $\hat{\mathbf{e}}_s \neq \mathbf{0}_k$, $\hat{\mathbf{b}}_s \to \hat{\mathbf{e}}_s$ for $\epsilon_H \to 0$ as $(\epsilon_H \hat{H}_o(j,i)\hat{b}_g(j))/\hat{e}_s(j) \to 0$.

If $k > 0$, the final beliefs are $\hat{\mathbf{b}}_s = \epsilon_H \hat{\mathbf{H}}_o \sum_{t \in N(s)} w_{ts}\hat{\mathbf{b}}_t$, i.e. the sum of all weighted neighbor's beliefs, transformed by $\epsilon_H \hat{\mathbf{H}}_o$. It follows that those final beliefs are by at least an order of $\epsilon_H$ smaller than those of nodes with explicit beliefs. It follows that for a node $e$ with geodesic number $k = 1$, $\hat{\mathbf{b}}_s \to \epsilon_H \hat{\mathbf{H}}_o \sum_{t \in N_0(e)} w_{ts}\hat{\mathbf{b}}_t$ where $N_0(e)$ is the subset of neighbors with explicit beliefs.

In turn, the propagated influence of this node to any other neighbor $u$ is $w_{su}\epsilon_H^2\hat{\mathbf{H}}_o^2 \sum_{t \in N_0(e)} w_{ts}\hat{\mathbf{b}}_t$. It follows by induction that the influence of a node $g_1$ with geodesic number $k_1$ on its neighbors is of an order of $\epsilon_H^{k_1-k_2}$ smaller than a node node $g_2$ with with geodesic number $k_2$ and $k_1 > k_2$, assuming that weights in the order of 1. It further follows from induction that for a node $e$ with $k > 0$, $\hat{\mathbf{b}}_s \to \epsilon \hat{\mathbf{H}}_o \sum_{t \in N_{k-1}(e)} w_{ts} \hat{\mathbf{b}}_t$ where $N_{k-1}(e)$ is the subset of neighbors with geodesic number $k-1$. It follows that the final beliefs are calculated inductively by traversing edges to nodes with smaller geodesic numbers back to explicit beliefs. Therefore, all such paths leading to explicit beliefs have the same length $k$. Let $P_s^k$ stand for the set of such paths. Then,

$$\hat{\mathbf{b}}_s \to \epsilon_H^k \hat{\mathbf{H}}_o^k \sum_{p \in P_s^k} w_p \hat{\mathbf{e}}_p \tag{31}$$

Since standardization of $\hat{\mathbf{b}}_s$ is equivalent to dividing it by its standard deviation, we see that $\hat{\mathbf{b}}'_s = \zeta(\hat{\mathbf{b}}_s)$ converges towards the same assignment as SBP for $\epsilon_H \to 0$. □

**Proposition 21: Initial assignment in SBP.**

PROOF PROPOSITION 21. There are two types of edges $(s \to t)$: "*valid*" edges where $g(t) = g(s) + 1$, i.e. edges that propagate beliefs from nodes $s$ to $t$, and "*invalid*" edges where $g(t) \leq g(s)$, i.e. directed edges where the target has a smaller or equal geodesic number and hence does not receive beliefs along this edge.

At iteration $i$, the algorithm calculates the implicit beliefs for all and only those nodes $v$ with $g(v) = i$. This follows from induction on the fact that in iteration $i$, tables $G$ and $B$ contain information for all nodes $v$ with $g(v) \leq i - 1$. Thus, by end of iteration $i$, all nodes with $g(v) \leq i$ are correctly calculated from valid edges only.

Finally, there cannot be any node $v$ with $g(v) = i$ if there is no other node $v'$ with $g(v') = i - 1$. Since the longest self-avoiding path in a graph is finite, the algorithm terminates in a finite number of iterations with the correct belief assignment. If there are nodes in the graph missing from $B$ at the end of the algorithm, then the graph must have distinct strongly connected components and those nodes are not connected from any node with explicit beliefs. □

**Proposition 22: Update beliefs in SBP.**

PROOF PROPOSITION 22. We need to show that all nodes which are affected by inserted explicit beliefs are correctly updated when Algorithm 3 terminates. There are two types of updated nodes: (1) nodes that keep their geodesic number $g'(v) = g(v)$, and which receive their implicit beliefs from their original source, but now together with new sources; and (2) nodes whose number decreases $g'(v) < g(v)$, and whose beliefs are now received only by new nodes (the number can never increase as inserts of new nodes never remove an existing shortest path). Note that for both types, there need to be a new explicit node that was inserted at exactly $g'(v)$ hops away.

We now show by induction that at iteration $i$, the algorithm calculates the implicit beliefs for all and only those nodes $v$ with $g'(v) = i$. We need to show two things: (1) that at iteration $i$, each node $v$ with $g'(v) = i$ is identified in table $G_n$, and (2) that the beliefs in table $B_n$ are correctly calculated. For (1) note that nodes to be updated at iteration $i$ are identified by following edges from nodes updated at iteration $i-1$ and only ignoring those nodes $v$ that already have $g'(v) < i$. By induction, by the end of iteration $i-1$ all such nodes are identified. For (2) note that calculating the updated beliefs, all parents $s$ with $g'(s) = i - 1$ are

---

**Algorithm 4:** ($\Delta$SBP:newEdges) Updates $B$ and $G$, given new weighted edges $A_n$ and original weighted network $A$.

**Input**: $A_n(s, t, w)$, $A(s, t, w)$
**Output**: Updated $B(v, c, b)$ and $G(v, g)$

1  Update main adjacency matrix
   $!A(s, t, w) :\!- A_n(s, t, w)$
2  Update geodesic numbers for seed nodes:
   $G_n(t, min(g_s+1)) :\!- G(s, g_s), A_n(s, t, \_), \neg(G(t, g_t), g_t < g_s)$
   $!G(v, g) :\!- G_n(v, g)$
3  Update beliefs for seed nodes:
   $B_n(t, c_2, sum(w \cdot b \cdot h)) :\!- G_n(t, g), A(s, t, w), B(s, c_1, b),$
   $\qquad\qquad G(s, g-1), H(c_1, c_2, h)$
   $!B(v, c, b) :\!- B_n(v, c, b)$
4  **repeat**
5  | Find next nodes to update:
   | $G'_n(t, min(g_s+1)) :\!- G_n(s, g_s), A(s, t, \_), \neg(G(t, g_t), g_t < g_s)$
   | $!G(v, g) :\!- G'_n(v, g)$
6  | Calculate new beliefs for these nodes:
   | $B_n(t, c_2, sum(w \cdot b \cdot h)) :\!- G'_n(t, g), A(s, t, w), B(s, c_1, b),$
   | $\qquad\qquad G(s, g-1), H(c_1, c_2, h)$
   | $!B(v, c, b) :\!- B'_n(v, c, b)$
7  | Delete $G_n$ and rename $G'_n$ to $G_n$
8  **until** no more inserts into $G'_n$
9  **return** $B$ and $G$

---

used together with their updated beliefs. By induction, all those beliefs were correctly calculated by the end of iteration $i - 1$. □

## C. $\Delta$SBP: ADDITION OF EDGES

Algorithm 4 gives the SQL translation for batch inserts of additional edges assuming a new table $A_n(\underline{s, t}, w)$ containing the set of new edges.

Note that this algorithm is more intricate than Algorithm 3 for the batch insert of explicit beliefs because edges and nodes can now be *visited more than once*. We provide here some intuition: In both Algorithm 2 and Algorithm 3, we started from original or new explicit beliefs, and were progressing the information in each iteration along paths of increasing geodesic numbers. This guaranteed that every node is updated only once as paths from later iterations to a node cannot contain shorter geodesic paths, and can thus ignored. In the case of added edges, however, we now have "seed nodes" from which updates need to be propagated throughout the immediate neighborhoods that have *different geodesic numbers*. Since these seed nodes can start propagating with different geodesic numbers in the first iteration, there may be later paths arriving at a node that actually have a shorter geodesic length than the previously arriving one. There are two cases: (1) if both nodes ending at a newly inserted edge have the same geodesic numbers before the update, then this edge cannot contain a geodesic paths and can thus be ignored; (2) if one node $s$ has a smaller geodesic number than the other node $t$ ($g_s < g_t$), then the direction $s \to t$ contains a new geodesic path, and $t$ becomes a new seed node. If (2a) $g_s + 1 = g_t$ before the update, then the new edge contains only one additional geodesic path for $t$ and $g_t$ does not change. If, however, (2b) $g_s + 1 < g_t$, then the new edge contains all geodesic paths for $t$, whose new geodesic number becomes $g'_t = g_s + 1$.

Line 1 updates the adjacency matrix. Line 2 then finds the first set of nodes that need their beliefs updated; those are the ones $t$ that are the target of any new edge $s \to t$ for which the geodesic number of the source is smaller than the target ($g_s < g_t$), even before the update. Note that several edges may have been added to the same target, and we thus have to use the minimum of the newly found geodesic paths. Line 3 then updates beliefs of nodes $t$ by following only edges from source nodes $s'$ with geodesic number $g_{s'} = g_t - 1$. In each subsequent iteration (line 4), we now keep two tables, $G_n(\underline{v}, g)$ for the nodes that were updated in the previous iteration, and $G'_n(\underline{v}, g)$ for the nodes to be updated during this iteration. Finding the new nodes to update $G'_n$ (line 5) and updating the beliefs of those nodes $B'_n$ (line 6) are similar as in the first iteration, except for the renaming of the tables (line 7). The algorithm stops when there are no more nodes to update (line 8).

| (a) Equation 20. | (b) Top belief assignment. | (c) Algorithm 2 (line 4 for $i\!=\!1$). | (d) Equation 32. |
|---|---|---|---|
| create table H2 as<br>select H1.c1, H2.c2,<br>   sum(H1.h*H2.h) as h<br>from H H1, H H2<br>where H1.c2 = H2.c1<br>group by H1.c1, H2.c2 | (select B.v, B.c<br>from B,<br>  (select B2.v, max(B2.b) as b<br>  from B B2<br>  group by B2.v) as X<br>where B.v = X.v and B.b = X.b) | insert into G<br>(select A.s, '1' from G, A<br>where G.v = A.s<br>and G.g = '0'<br>and A.d not in<br>  (select G.v from G)) | delete from B<br>where v in<br>  (select Bn.v from Bn);<br>insert into B<br>select * from Bn; |

Figure 9: SQL for example Datalog statements from Sect. 5.3 and Sect. 6.3.

Recall, that the geodesic paths and the beliefs of a node may be updated more than once. For this reason, some care needs to be taken when updating the result table $B_n$. For example, consider two new edges $s - v$ and $v - t$ where the adjacent nodes have original geodesic numbers $g_{s(0)} = 0$, $g_{v(0)} = 2$, and $g_{t(0)} = 4$. Then, both $v$ and $t$ become seed nodes for the first iteration with new geodesic numbers $g_{v(1)} = 1$, and $g_{t(1)} = 3$. Note since updates happen concurrently $t$ still inherits $g_{t(1)} = g_{v(0)} + 1$. Beliefs are not updated during this iteration. In the iteration 2, the updated information travels from $v$ to $t$ and we have $g_{t(2)} = 2$. Using similar ideas, one can create a pathological example, in which updating of the graph takes $n^2$ time with our algorithm, where $n$ is the number of added edges.

We next sketch an idea that would avoid the quadratic time increase: sort the seed nodes by new updated geodesic numbers and let $i$ be the smallest updated geodesic number across all seed nodes. Start propagating beliefs, in each iteration $j$, from seed nodes with new geodesic number $i + j - 1$, unless this nodes was updated in a previous iteration. This simple change would avoid that a node gets updated its beliefs more than once. We have not implemented this idea and leave experimenting with it for future work.

PROPOSITION 24 (ALGORITHM 4). *Algorithm 4 terminates in finite number of iterations and returns a sound and complete enumeration of updated beliefs.*

PROOF PROPOSITION 24. The proof for Proposition 24 works similar to the one for Proposition 22, with three differences: (1) the induction base are those nodes which were the target of a newly inserted edge and whose beliefs thus need to be updated ($g_s < g_t$); (2) we need to use a grouping and minimum when updating the geodesic number; and (3) the induction is now on on the distance from the original "seed nodes." Notice that now the geodesic number and the beliefs of a node can change more than once, which is taken care of our algorithm by maintaining two separate tables $G'_n$ and $G_n$. □

## D. EXAMPLE SQL CODE

We give here two examples for SQL translations from our Datalog notation in Sect. 5.3 and Sect. 6.3: Figure 9a gives the SQL equivalent for the Datalog statement in Eq. 20. Figure 9b returns the top beliefs for each node. Figure 9c shows the query from line 4 representing negation on relational atoms with anonymous variables.

Figure 9d shows the SQL equivalent for the following update that uses an exclamation mark left of a Datalog query to imply that the respective data record is either inserted or an existing one updated:

$$!B(v, c, b) := B_n(v, c, b) \qquad (32)$$

## E. ONE CLASS ($K = 2$)

Previous work [25] has given a linearization for belief propagation for the binary case ($k = 2$). Here we show that our more general results include the binary case as special case.

We start from Eq. 27 and use the normalization conditions for $k = 2$ to write $\hat{\mathbf{b}} = \begin{bmatrix} \hat{b} \\ -\hat{b} \end{bmatrix}$, $\hat{\mathbf{m}} = \begin{bmatrix} \hat{m} \\ -\hat{m} \end{bmatrix}$, and $\hat{\mathbf{H}} = \begin{bmatrix} \hat{h} & -\hat{h} \\ -\hat{h} & \hat{h} \end{bmatrix}$. We then get $\hat{\mathbf{H}}^2 = 2 \begin{bmatrix} \hat{h}^2 & -\hat{h}^2 \\ -\hat{h}^2 & \hat{h}^2 \end{bmatrix}$. As the normalization $\mathbf{x}(1) = -\mathbf{x}(1)$ holds for all results, it suffices to only focus on one dimensions, which we choose w.l.o.g. to be the first. We get: $(k\hat{\mathbf{H}}\hat{\mathbf{b}}_s)(1) = 4\hat{b}_s\hat{h}$, $(k\hat{\mathbf{H}}^2\hat{\mathbf{b}}_t)(1) = 8\hat{b}_t\hat{h}^2$, $(\mathbf{I}_k - \hat{\mathbf{H}})(1) = 1 - 4\hat{h}^2$, and finally:

$$\hat{m}_{s,t} = \frac{4\hat{h}}{1 - 4\hat{h}^2}\hat{b}_s - \frac{8\hat{h}^2}{1 - 4\hat{h}^2}\hat{b}_t \qquad (33)$$

Note that our Eq. 33 differs from [25, Eq. 19] by a factor of 2 on the right side. This is result of the decision to center the messages around $\frac{1}{2}$ in [25], whereas this work decided to centered around 1: Centering messages around 1 allowed us to ignore incoming messages that have no residuals (i.e. $\mathbf{m}_{st} = \mathbf{1}$) and was a crucial assumption in our derivations (see Sect. 4.1). This difference leads to a factor 2 in Eq. 9 for $k = 2$, and thus a factor 2 difference in Eq. 33. However, both alternative centering approaches ultimately lead to the same equation in the binary case:

$$\hat{\mathbf{b}} = \left( \mathbf{I}_n - \frac{2\hat{h}}{1 - 4\hat{h}^2}\mathbf{A} + \frac{4\hat{h}^2}{1 - 4\hat{h}^2}\mathbf{D} \right)^{-1}\hat{\mathbf{e}}$$

where $\mathbf{b}$ and $\mathbf{e}$ are the column vectors that contain the first dimension of the binary centered beliefs for each node. This can be easily seen by writing the original, non-simplified version Eq. 29 in the vectorized form of Eq. 11.

Further note that experiments in [25] showed a decreasing quality of classification for decreasing $\epsilon_H$ and, thus, recommended $\epsilon_H$ to be chosen above a certain threshold. The conclusion from the present paper is that, in theory, the quality should converge quickly, and then remain constant

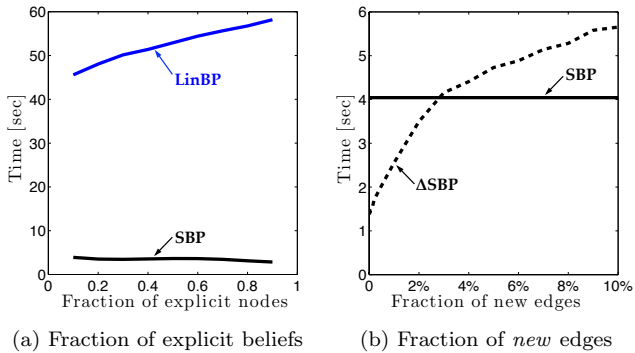

(a) Fraction of explicit beliefs

(b) Fraction of *new* edges

Figure 10: Additional scalability experiments in SQL: (a) SBP gets slightly faster with increasing fraction of explicit beliefs, in contrast to LinBP. (b,c): It is faster to update SBP when less than $\approx 60\%$ of the final explicit beliefs are new or less than $\approx 3\%$ of the final edges.

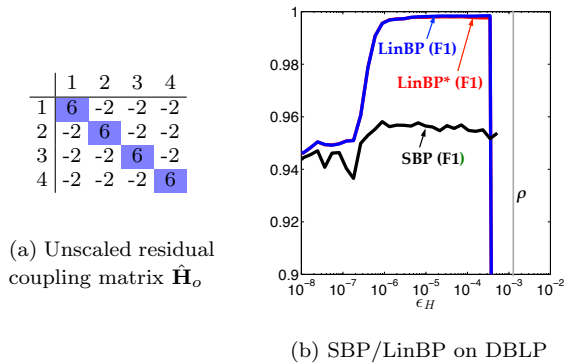

(a) Unscaled residual coupling matrix $\hat{\mathbf{H}}_o$

(b) SBP/LinBP on DBLP

Figure 11: Real world data used for our experiments.

for $\epsilon_H \to 0^+$. Any deviations from this behavior must be the consequences of unavoidable roundoff errors due to limited precision of floating-point computations. Our new semantics SBP captures the exact theoretical behavior for $\epsilon_H \to 0^+$ while avoiding the problem of roundoff errors.

## F. ADDITIONAL EXPERIMENTS

### F.1 Kronecker graphs

Here, we add two more experiments on our example Kronecker graphs to answer two more questions:

QUESTION 5. *Timing: How does the fraction of explicit beliefs in a fixed graph affect the time for LinBP and SBP?*

*Result 5.* LinBP gets slightly slower, while SBP get slightly faster with an increasing fraction of explicit beliefs.

Figure 10a illustrates the result on our Kronecker graph #5 using 5 iterations for LinBP. LinBP gets slightly slower because the LinBP update equations see an increased calculation load. In contrast, SBP gets slightly faster as the number of edges that propagate information decreases with a denser fraction of explicit beliefs. However, both effects are minor.

QUESTION 6. *Timing: When is it faster to update a graph incrementally than to recalculate from scratch with SBP?*

*Result 6.* In our experiments, it was faster to update SBP when less than $\approx 3\%$ of the final edges are new.

Figure 10b shows these results for SQL on graph #5. We keep 10% of explicit beliefs fixed, and vary a certain fraction of edges as *new* edges. For example, the 5% on the horizontal axis implies that we had 95% of all edges in the graph before the update, and are adding the remaining 5% of edges with the incremental Algorithm 4 ("$\Delta$SBP$_e$"). We show here only the first 10%; beyond this, the updates need increasingly more time until until up to $\approx 16$ seconds (4 times the necessary time to update from scratch). Thus, both update algorithms (updating beliefs or edges) can speed up the calculations. However, the range for which incremental updates are beneficial is smaller for edges than for beliefs.

This result seems to confirm our intuition that updating the graph structure is more complex than updating implicit beliefs. Future work is needed to see if the incremental edge updates can be made faster, in particular with the idea mentioned at the end of Appendix C.

### F.2 Experiment on DBLP data

For this experiment, we use the DBLP data set from [20] which consists of 36 138 nodes representing papers, authors, conferences, and terms. Each paper is connected to its authors, the conference in which it appeared and the terms in its title. Overall, the graph contains 341 564 edges (counting edges twice according to their direction). Only 3 750 nodes (i.e. $\approx 10.4\%$) are labeled explicitly with one of 4 classes: AI (Artificial Intelligence), DB (Databases), DM (Data Mining), and IR (Information Retrieval). We are assuming homophily, which is represented by the $4 \times 4$-matrix in Fig. 11a. Our goal is to label the remaining 89.6% of the nodes. We use the same experimental setup from Sect. 7 for determining accuracy of our methods against BP as GT. The F1-score corresponds to the harmonic mean of precision and recall. We again vary $\epsilon_H$ to get different heterophily matrices by using $\hat{\mathbf{H}} = \epsilon_H \hat{\mathbf{H}}_o$.

Figure 11b shows that LinBP and LinBP$^*$ approximate BP very well as long as BP converges. LinBP converges until $\epsilon_H \approx 0.0013$, however BP stops converging earlier: This explains the gap between when the accuracy drops and the actual convergence bounds for LinBP. On the left side, we see results from floating-point rounding errors. SBP performs worse than LinBP in this case due to many ties. The absolute accuracy, however, is still above 95%.

## G. COMPARISON WITH EXISTING CONVERGENCE BOUNDS FOR BP

Convergence of BP in loopy graphs has been studied in various works before [8, 19, 32]. To the best of our knowledge, all existing bounds for BP give only sufficient convergence conditions. In contrast, our work presents a stronger result by providing sufficient *and necessary conditions* for the convergence of LinBP (Eq. 16 and Eq. 17).

In the following, we compare our convergence bounds with one of the tightest *sufficient* bounds for the convergence of standard BP, the one proposed by Mooij et al. [32] (the same bound for the special case of pairwise, strictly positive potentials has been simultaneously derived by [19]). Similar to our Eq. 16 and Eq. 17, the bound in [32] considers the spectral radius of an appropriate matrix to study convergence. While we study the adjacency matrix $\mathbf{A}$ of the underlying network, [32] uses a special "edge matrix" $\mathbf{A}_{\text{edge}}$ which represents all *directed edges* in a network (i.e. $(u,v)$ and $(v,u)$ are treated separately). Writing $|E|$ for the number of edges in the network, $\mathbf{A}_{\text{edge}}$ is thus a $2|E| \times 2|E|$ matrix, where an edge $(u,v)$ is connected to all edges $(w,u)$ with $v \neq w$.

For our scenario of pairwise potentials and a single $\mathbf{H}$ matrix, the bound of [32] can be written as

$$c(\mathbf{H}) \cdot \rho(\mathbf{A}_{\text{edge}}) < 1 \qquad (34)$$

where $c(\mathbf{H})$ is a constant depending on the (non-centered) $\mathbf{H}$ matrix and given by

$$c(\mathbf{H}) := \max_{c_1 \neq c_2} \max_{d_1 \neq d_2} \tanh\left(\frac{1}{4} \cdot \left(\log \frac{\mathbf{H}_{c_1,d_1} \mathbf{H}_{c_1,d_2}}{\mathbf{H}_{c_2,d_1} \mathbf{H}_{c_2,d_2}}\right)\right)$$

By comparing Eq. 34 with our convergence bound for LinBP*

$$\rho(\hat{\mathbf{H}}) \cdot \rho(\mathbf{A}) < 1 \qquad \text{(cf. Eq. 17)}$$

we can draw two interesting comparisons:

1. Based on empirical analysis, it holds that $\rho(\mathbf{A}_{\text{edge}}) + 1 \approx \rho(\mathbf{A})$. Thus, in particular $\rho(\mathbf{A}_{\text{edge}}) < \rho(\mathbf{A})$. In other words, there are graphs $\mathbf{A}$ and heterophily matrices $\mathbf{H}/\hat{\mathbf{H}}$ for which Eq. 34 guarantees that BP converges, whereas LinBP does not.
2. In a multiple class setting, in contrast, we generally observe $c(\mathbf{H}) > \rho(\hat{\mathbf{H}})$. In other words, there are settings for which LinBP converges, while Eq. 34 does not guarantee convergence for BP.

Thus, we see that neither of the two bounds (Eq. 34 from [32] for BP, and our Eq. 17 for LinBP) subsumes the other, and neither of the two bounds guarantees a wider range of applicability of $\hat{\mathbf{H}}$ for their respective algorithm. However, for our use case of network data and multiple classes, Eq. 16 and Eq. 17 usually provide better results, i.e. they guarantee convergence for a larger set of $\hat{\mathbf{H}}$ matrices. The reason is that we usually have high-degree nodes in real networks; therefore, the spectral radii of $\mathbf{A}_{\text{edge}}$ and $\mathbf{A}$ become large and thus their difference vanishes whereas often $c(\mathbf{H}) > \rho(\hat{\mathbf{H}})$.

18